\documentclass[final]{jfm}
\usepackage{graphicx}
\usepackage[dvipsnames]{xcolor}
\usepackage[utf8]{inputenc}
\usepackage{amsmath,amssymb,amsfonts}
\usepackage{pict2e}
\usepackage{tikz}
\usepackage[longnamesfirst]{natbib}
\usepackage{parskip}
\paperheight\pdfpageheight
\paperwidth\pdfpagewidth
\usepackage[unicode,colorlinks,linkcolor=blue,citecolor=blue,urlcolor=blue]{hyperref}
\graphicspath{ {images/} }
\usetikzlibrary{patterns}

\newcommand{\intd}{\mathrm{d}} 
\newcommand{\I}{\mathrm{i}} 
\newcommand{\lr}[1]{\left( #1 \right)}
\newcommand{\pfrac}[2]{\frac{\partial #1 }{\partial #2 }}
\newcommand{\Dfrac}[2]{\frac{D #1 }{D #2 }}
\newcommand{\tp}{\widetilde{p}}

\newcommand{\bbk}{\lr{ \frac{\omega^2}{M^2} + \frac{m^2}{r_c^{+2}} }}
\newcommand{\bk}{\frac{\omega}{M}}
\newcommand{\tG}{\widetilde{G}}
\newcommand{\rg}{\hat{r}}
\renewcommand{\Imag}{\mathrm{Im}}
\renewcommand{\Real}{\mathrm{Re}}
\newcommand{\eps}{\varepsilon}
\newcommand{\rl}{\check{r}}
\newcommand{\cg}{\widehat{C}}
\newcommand{\cl}{\reallywidecheck{C}}
\newcommand{\dg}{\widehat{D}}
\newcommand{\dl}{\reallywidecheck{D}}
\newcommand{\mi}{\I}
\newcommand{\rmatch}{\Bar{r}}
\newcommand{\ex}{\mathrm{e}}

\usepackage{scalerel,stackengine}
\stackMath
\newcommand\reallywidecheck[1]{%
\savestack{\tmpbox}{\stretchto{%
  \scaleto{%
    \scalerel*[\widthof{\ensuremath{#1}}]{\kern-.6pt\bigwedge\kern-.6pt}%
    {\rule[-\textheight/2]{1ex}{\textheight}}
  }{\textheight}%
}{0.5ex}}%
\stackon[1pt]{#1}{\scalebox{-1}{\tmpbox}}%
}

\newcommand{\set}[1]{#1}
\newcommand{\setA}{\set{A1}}
\newcommand{\setB}{\set{A2}}

\newcommand{\setD}{\set{B}}
\newcommand{\setE}{\set{C}}

\def\clap#1{\hbox to 0pt{\hss#1\hss}}

\def\mathrlap{\mathpalette\mathrlapinternal}
\def\mathclap{\mathpalette\mathclapinternal}

\def\mathrlapinternal#1#2{%
\rlap{$\mathsurround=0pt#1{#2}$}}
\def\mathclapinternal#1#2{%
\clap{$\mathsurround=0pt#1{#2}$}}

\shorttitle{The Critical Layer in Parabolic Boundary Layers}
\title[The critical layer in quadratic boundary layers]{The critical layer in quadratic flow boundary layers over acoustic linings}
\shortauthor{King, Brambley, Liupekevicius, Radia, Lafourcade and Shah}
\author{Matthew~J.~King\aff{1}
\and Edward~J.~Brambley\aff{1,2}%
\corresp{\email{E.J.Brambley@warwick.ac.uk}}\\
\and Renan~Liupekevicius\aff{2,3}
\and Miren~Radia\aff{4}
\and Paul~Lafourcade\aff{5,6}
\and Tauqeer~H.~Shah\aff{7,8}%
}
\affiliation{%
\aff{1}Mathematics~Institute, University~of~Warwick, Coventry~CV4~7AL, UK
\vspace{-2ex}%
\aff{2}WMG, University~of~Warwick, Coventry~CV4~7AL, UK
\vspace{-2ex}%
\aff{3}Mechanical~Engineering, Eindhoven~University~of~Technology, 5600~MB~Eindhoven, NL
\vspace{-2ex}%
\aff{4}DAMTP, University~of~Cambridge, Cambridge~CB3~0WA, UK
\vspace{-2ex}%
\aff{5}CEA~DAM~\^Ile-de-France, 91297~Arpajon, France
\vspace{-2ex}%
\aff{6}LMCE, CEA~Paris-Saclay, 91680~Bruy\`eres-le-Ch\^atel, France
\vspace{-2ex}%
\aff{7}Department~of~Physics, Government~College~University~Faisalabad, Faisalabad, Pakistan
\vspace{-2ex}%
\aff{8}Faculty~of~Technology, Linnaeus~University, 351~95~V\"axj\"o, Sweden
}
\date{19~December~2021; revised~28~June~2022; accepted~25~August~2022.}
\pubyear{2022}
\volume{950}
\pagerange{A8}
\doi{jfm.2022.753}
\copyrightnotice{\copyright\ The authors, 2022. This work is licensed under a Creative Commons Attribution (CC BY) licence (\url{http://creativecommons.org/licenses/by/4.0}).}

\begin{document}

\maketitle

{\centering This work has been published in the Journal of Fluid Mechanics.  The official version (the ``Version of Record'') is available from the journal website (\url{https://doi.org/10.1017/jfm.2022.753}).\par}

\begin{abstract}
A straight cylindrical duct is considered containing an axial mean flow that is uniform everywhere except within a boundary layer near the wall, which need not be thin.  Within this boundary layer the mean flow varies parabolically.  The linearized Euler equations are Fourier transformed to give the Pridmore-Brown equation, for which the Greens function is constructed using Frobenius series.  Inverting the spatial Fourier transform, the critical layer contribution is given as the non-modal contribution from integrating around the continuous spectrum branch cut. This contribution is found to be the dominant downstream contribution to the pressure perturbation in certain cases, particularly for thicker boundary layers.  The continuous spectrum branch cut is also found to stabilize what are otherwise convectively unstable modes by hiding them behind the branch cut.  Overall, the contribution from the critical layer is found to give a neutrally stable non-modal wave with a phase velocity equal to the mean flow velocity at the source when the source is located within the sheared-flow region, and to decay algebraically along the duct as $O(x^{-\frac{5}{2}})$ for a source located with the uniform flow region.  The Frobenius expansion, in addition to being numerically accurate close to the critical layer where other numerical methods loose accuracy, is also able to locate modal poles hidden behind the branch cut, which other methods are unable to find; this includes the stabilized hydrodynamic instability.  Matlab code is provided to compute the Greens function.

\vspace{1em}
\noindent{\bf Key words:}\hspace{1em} aeroacoustics, boundary layer stability, compressible boundary layers
\end{abstract}

\section{Introduction}
The propagation of sound through an otherwise steady mean flow has many important applications.  One such application is predicting and optimizing aircraft engines noise.  With aircraft noise being subjected to ever increasing restrictions, being able to successfully model this noise becomes increasingly important.  In particular, aircraft engine noise at takeoff depends critically on the sound absorbing performance of acoustic liners.  Unfortunately, acoustic liner performance in the presence of a steady mean flow is poorly predicted by existing theory, as demonstrated by comparisons to laboratory experiments~\citep{renou2011failure,spillere2020experimentally}.  The theory is equally applicable to any situation with small perturbations to an otherwise steady mean flow along a non-rigid boundary: for example, the stability analysis of flow over a deformable surface.

The behaviour of sound in an otherwise steady mean flow is usually modelled using the linearized Euler Equations.  Non-rigid boundaries, such as the acoustic liners used in aircraft engines, are usually modelled using an impedance boundary condition, where a disturbance with oscillating pressure $\Real(p\exp\{\I\omega t\})$ leads to an oscillating normal boundary velocity $\Real(v\exp\{\I\omega t\})$ given by $p = Z(\omega)v$.  Such impedance boundary conditions are well understood for a mean flow that satisfies no-slip at the boundary.  Often, however, we use a simplified model where the mean flow does not satisfy no-slip at the boundary: for example, uniform axial flow in a duct.  For slipping mean flows, it is known that the impedance boundary condition must be modified.  A common modified boundary condition is the Myers, or Ingard--Myers, boundary condition~\citep{ingard1959influence,myers1980acoustic}.  This boundary condition is known to be the correct limiting behaviour for an inviscid mean flow boundary layer in the limit that the boundary layer thickness tends to zero~\citep{eversman1972transmission,tester1973some}.  However, this boundary condition, when applied in the time domain, is ill-posed~\citep{brambley2009fundamental}.  Several alternative boundary conditions have been suggested~\citep{brambley2011well,schulz2017momentumtransfer,khamis2017twodeck,auregan2018stress}, which each attempt to include more relevant physics, including the effect of the mean flow boundary layer and the effect of viscosity.  However, these boundary conditions come with their own complications, including the need to fit further free parameters, and as yet none have been made to agree with laboratory experiments~\citep{spillere2020experimentally}.

In light of this difficulty with boundary conditions in slipping mean flow, one may instead only consider mean flows $U(r)$ that satisfy no-slip at the boundary~\citep[e.g.][]{weng2017flow}.  Doing so, however, involves solving for the sound in a strongly varying mean flow, which is especially taxing when the boundary layers are particularly thin.  Numerically resolving the sound in thin boundary layers requires a fine resolution, which then also requires a small timestep owing to the CFL condition.  Progress may be made analytically by considering the simplified situation of a straight rectilinear or cylindrical duct containing axial mean flow (as depicted later in figure~\ref{figCyl}).  By linearizing the Euler Equations about this steady mean flow and assuming $\exp\{\I\omega t - \I kx\}$ dependence, one eventually arrives at the Pridmore-Brown equation~\eqref{PridmoreBrownFull}, a second-order linear ODE for the pressure perturbation within the duct due to~\citet{pridmore1958sound}.  The Pridmore-Brown equation has been the subjected of much analysis~\citep[e.g.][]{mungur1969acoustic,ko1972sound,swinbanks1975sound,nagel1982boundary,brambley2012critical,rienstra2020numerical}, owing to its complexity.  One complexity is that, treating the frequency $\omega$ as known and solving for the axial wavenumber $k$ as the eigenvalue, the Pridmore-Brown equation is not Sturm--Liouville and results in a nonlinear eigenvalue problem for $k$.  A second complexity is that the Pridmore-Brown equation possesses a regular singularity, referred to as a \emph{critical layer} or \emph{continuous spectrum}.  Despite these difficulties, eigenfunction expansions using eigenfunctions of the Pridmore-Brown equation are frequently used, with the eigenfunctions assumed to form a complete basis (despite the problem being non-self-adjoint) and the effect of the critical layer ignored~\citep[e.g.][]{brooks2007sound,olivieri2010determining,oppeneer2016efficient,rienstra2021slowly}.

The lack of completeness of the modal solutions of the Pridmore-Brown equation motivates the investigation of the Green's function solution.  The Green's function is the solution of the governing equations subject to a point forcing; for example, a point mass source leads to the right-hand-side of equation~\eqref{PridmoreBrownFull}.  The Green's function may be used to construct the solution of the governing equations subject to any arbitrary forcing; hence, the Green's function is capable of being used to express any solution to the governing equations, in contrast to a modal eigenvalue expansion which can only express an arbitrary solution if the modal basis is complete.  The Green's function is also worth considering on its own merits without reference to a particular forcing, since if the governing equations are capable of exhibiting a particular feature (such as instability, focusing, perfect reflection, etc), then the Green's function must also exhibit that feature.  The Green's function is also used in various approximation techniques~\citep[e.g.][]{brambley2012eigenmodes,posson2013acoustic,mathews2018serrated}.  For this reason, the Green's functions has been constructed for a variety of acoustical situations~\citep[e.g.][]{rienstra2008greens,brambley2012critical,mathews2017greens,mathews2018entropy}.  In particular, the Green's function solution to the Pridmore-Brown equation naturally includes the critical layer. 

The critical layer, or continuous spectrum, is a singularity of the linearized Euler equations occurring when the phase velocity of the perturbation, $\omega/k$, is equal to the local fluid velocity of the steady flow, $U(r_c)$, for some critical radius $r_c$.  Because the phase speed is equal to the flow speed, the effect of the critical layer may be thought of as being convected with the mean flow, and therefore as hydrodynamic in nature~\citep{case1960stability,rienstra2013trailing}.  For swirling flows, the critical layer is known to lead to algebraically growing instabilities~\citep{golubev1996sound,tam1998wave,heaton2006algebraic}.  For the Pridmore-Brown equation, the critical layer is currently thought to lead to algebraically decaying disturbances, although publications differ on the exact nature of the decay.  For example, \citet{swinbanks1975sound} predicted a disturbance of constant amplitude plus a disturbance with $O(x^{-3})$ decay for a point source, and $O(x^{-1})$ decay for a distributed source, although exact formulae for these disturbances are not given.  \Citet[][p.~62]{swinbanks1975sound} goes on to argue that the constant amplitude disturbance would not be present when the disturbance is caused ``by moving the surface of a solid body''.  In contrast, \citet{felix2007acoustic} demonstrated numerically, for a point source in a parabolic mean flow, a decay rate of $O(x^{-1})$.  More recently, \citet{brambley2012critical} gave an explicit analytic solution for the critical layer far-field response for a mean flow $U(r)$ that is constant in the centre of the duct, and then varies linearly in a ``boundary layer'' region to zero at the duct walls.  Locating a point source at a radius $r_0$, they found the pressure perturbation from the critical layer at a radius $r$ consisted of three distinct components with phase velocities $U(0)$, $U(r)$ and $U(r_0)$, each with different decay rates. However, \citet{brambley2012critical} chose a rather special mean flow profile.  In particular, the critical layer is usually caused by a nonzero second derivative of the mean flow profile, $U''(r)$, but for the constant-then-linear mean flow $U''(r)$ is either identically zero or has a delta function discontinuity; in the constant-then-linear case, \citet{brambley2012critical} instead attributed the critical layer to the cylindrical geometry.

In many cases, the effect of the critical layer is negligible in comparison with the modal sum of the acoustics modes.  However, when all acoustic modes are cut-off and non-propagating, the effect of the critical layer will be dominant.  Moreover, \citet[][figure~6]{brambley2013surface} showed that a mode representing a hydrodynamic instability could interact with the critical layer, although this was not seen for a constant-then-linear mean flow profile.

Since the critical layer is a singularity of the Pridmore-Brown equation, traditional numerical methods are particularly inaccurate near the critical layer.  This often manifests as a collection of spurious numerical modes being located along the critical layer.  In contrast, previous studies have used a Frobenius expansion about the singular point $r=r_c$~\citep[e.g.][]{heaton2006algebraic,campos2009propagation,brambley2012critical}.  This technique both gives increasing accuracy as the critical layer is approached, and allows analytical continuation behind the critical layer branch cut.  For example, \citet[][figure~10]{brambley2012critical} found a previously unknown mode close to the critical layer that was unable to be resolved numerically using more traditional finite differences.  One complication of the Frobenius series, however, is that, much like a power series, it has an associated radius of convergence.  For the constant-then-linear mean flow Frobenius expansion~\citep{brambley2012critical}, this did not prove a problem, as the radius of convergence covered the region of interest in all cases that were considered.  For general flow profiles this will not be the case, and a solution covering the entire region of interest will involve multiple Frobenius expansions with overlapping radii of convergence; this will turn out to be the case here.  By matching two different expansions in a region where both converge, a hybrid solution may be constructed that spans the whole region of interest.

Here, we use the Frobenius expansion method as described by~\citet{brambley2012critical}, and apply it to a mean flow that is constant in the centre of the duct and then varies quadratically within a boundary layer to satisfy non-slip at the wall.  As well as being more realistic than the constant-then-linear profile considered by~\citet{brambley2012critical}, this mean flow profile is twice differentiable, allowing $U''(r)$ to enter the analysis, and as such we expect the results to be more representative of an arbitrary mean flow profile.  The Frobenius expansion is derived in section~\ref{section:formulation}, along with a derivation of the Pridmore-Brown equations by spatially Fourier transforming the linearized Euler equations.  The Frobenius expansion is then used in section~\ref{section:greens} to derive the Green's function for a point mass source, including inverting the spatial Fourier transform and investigating the far-field behaviour.  Results are presented in section~\ref{section:results} by numerically evaluating the Frobenius expansions and the Green's function.  These results are compared against previous results, particularly against the predictions by~\citet{swinbanks1975sound} and the constant-then-linear results by~\citet{brambley2012critical}.  Finally, the implications of this work are discussed, and areas for further research highlighted, in section~\ref{section:conclusion}.

\section{Problem Formulation and Homogeneous Solutions}
\label{section:formulation}

\subsection{Constructing the Pridmore-Brown Equation}
The governing equations for what follows are the Euler equations  with a mass source $q$,
\begin{align}\label{Euler}
    \pfrac{\rho}{t}+\boldsymbol{\nabla}\cdot(\rho \boldsymbol{u})&=q, & \rho\Dfrac{\boldsymbol{u}}{t}&=-\boldsymbol{\nabla}p,& \Dfrac{p}{t}&=c^2\Dfrac{\rho}{t},
\end{align}
where $\boldsymbol{u}$ is the fluid velocity, $p$ is the pressure, $\rho$ is the density, and $c^2 = \frac{\partial p}{\partial\rho}|_s$ is the square of the sound speed.  In what follows, we take the mass source $q$ to be a small time-harmonic point mass source.  In cylindrical coordinates $(x,r,\theta)$, with a suitable choice of origin, this mass source $q$ may in general be taken as
\begin{equation}
q = \Real\left(\frac{\epsilon}{r_0}\delta(x)\delta(\theta)\delta(r-r_0)\exp\{\I\omega t\}\right),
\end{equation}
where $\epsilon$ is the small amplitude, $\omega$ is the frequency, and the $1/r_0$ term comes from writing a unit amplitude point source in cylindrical coordinates.  We expand each variable in powers of $\epsilon$,
\begin{gather}
\rho=\rho_0(r)+\Real\big(\epsilon\hat{\rho}e^{\mi\omega t}\big) + O(\epsilon^2),
\qquad\qquad
     p=p_0+\Real\big(\epsilon \hat{p}e^{\mi\omega t}\big) + O(\epsilon^2),
\notag\\
\boldsymbol{u}=U(r)\boldsymbol{e_x} + \Real\Big(\epsilon\big(\hat{u}, \hat{v}, \hat{w} \big)e^{\mi\omega t}\Big) + O(\epsilon^2),
\qquad\qquad
c^2 = c_0^2(r) + O(\epsilon),
\end{gather}
where $p_0$ is necessarily a constant in order that the steady state should satisfy the Euler equations, and it turns out that $c^2$ is only needed to leading order in what follows.  Without loss of generality, all perturbations are expanded using a Fourier series in $\theta$ and a Fourier Transform in $x$.  As a result, the pressure perturbation is given as
\begin{equation}\label{fourierSumfourierInv}
    \hat{p}(x,r,\theta)=\frac{1}{2\upi} \sum_{m=-\infty}^\infty \ex^{-\mi m \theta} \int_{-\infty}^\infty \tp(r;k,m,\omega)\ex^{-\mi k x}\,\intd k,
\end{equation}
and similarly for the density $\hat{\rho}$ and the velocity components $\hat{u}$, $\hat{v}$ and $\hat{w}$.  Substituting these into the Euler equations~\eqref{Euler}, and linearizing by ignoring terms of $O(\epsilon^2)$ or smaller, each of $\tilde{\rho}$, $\tilde{u}$, $\tilde{w}$, and finally $\tilde{v}$ may be eliminated, to leave a second order ODE in the radial coordinate $r$ for $\tp$,
\begin{subequations}\label{PridmoreBrownFull}\begin{gather}\label{PridmoreBrown}
    \tp^{\prime\prime}+\lr{\frac{2k U^\prime}{\omega-U(r)k}+\frac{1}{r}-\frac{\rho^\prime_0}{\rho_0}}\!\tp^\prime+\lr{\frac{(\omega-U(r)k)^2}{c_0^2}-k^2-\frac{m^2}{r^2}}\!\tp=\frac{\omega-U(r_0)k}{2\I\upi r_0}\delta(r-r_0),
    \\
    \text{with}\qquad\qquad\tilde{v} = \frac{\I\tp'}{\rho_0(\omega - Uk)},
    \label{equ:vp}
\end{gather}\end{subequations}
where a prime denotes the derivative with respect to $r$.  This is the \citet{pridmore1958sound} equation for a point mass source, written in cylindrical co-ordinates.

One boundary condition to~\eqref{PridmoreBrownFull} is regularity at $r=0$.  The singular solution behaves, for $m\neq 0$, as $O(r^{-|m|})$ as $r\to 0$, and the regular solution behaves as $O(r^{|m|})$.  For $m=0$, the singular solution behaves as $O(\log r)$ while the regular solution behaves as $O(1)$.  Eliminating the singular solution is therefore possible using the boundary conditions at $r=0$
\begin{align}\label{r=0BndCon}
    \tp(0)&=0 \quad \text{for }m\not=0 & \tp^\prime(0)&=0 \quad \text{for }m=0
\end{align}
To model sound within a straight cylindrical duct of radius $r=a$, we take the other boundary condition to be the impedance boundary condition at $r=a$,
\begin{align}\label{r=1BndCon}
    \tp(a) &= Z(\omega)\tilde{v}(a)&
    &\iff&
    \tp^\prime(a)&=-\frac{\mi\omega\rho_0}{Z}\tp(a),
\end{align}
where $Z(\omega)$ is the impedance of the duct wall, and the two expressions are equivalent in light of~\eqref{equ:vp}. A hard wall corresponds to $Z \to \infty$, and hence to $\tilde{v}(a) = 0$, or equivalently to $\tilde{p}'(a) = 0$.

\begin{figure}
    \centering
\begin{tikzpicture}
\draw (-5,2) -- (7,2);
\draw (-5,-2) -- (7,-2);
\fill (-3.5,2.25) rectangle (5.5,2);
\draw (-1,2.23) -- (0,2.23) node[anchor= south ] {$Z=p/v$};
\fill (-3.5,-2) rectangle (5.5,-2.25);
\draw (0,-2) .. controls (1,-1.8) and (1,-1.6)  .. (1,-1.5);
\draw (0,2) .. controls (1,1.8) and (1,1.6)  .. (1,1.5);
\draw (1,1.5) -- (1,-1.5);
\draw [dashed] (0,2) -- (0,-2);
\draw[dashed] (3.5,2) -- (3.5,-2);
\draw[thick,->] (-3,0) -- (-2,0) node[anchor= west] {$x$};
\draw[thick,->] (-3,0) -- (-3,1.5) node[anchor= south] {$r$};
\draw[<->] (2,-2) -- (2,-1.5) node[anchor= north west] {$h$};
\draw[thick, ->] (0,0) -- (1,0) node[anchor= west] {$U(r)$};
\draw[thick, ->] (0,-0.5) -- (1,-0.5);
\draw[thick, ->] (0,1.5) -- (1,1.5);
\draw[thick, ->] (0,1) -- (1,1);
\draw[thick, ->] (0,0.5) -- (1,0.5);
\draw[thick, ->] (0,-1) -- (1,-1);
\draw[thick, ->] (0,-1.5) -- (1,-1.5);
\draw (3.5,2) -- (3.5,2) node[anchor= north west ] {$\rho_0(r)$};
 \draw[thick, ->] (-2.55,0.45) arc (30:330:0.5cm and 1cm) node[anchor= west] {$\theta$};
\end{tikzpicture}
    \caption{A cross sectional view of a cylindrical duct with lined walls containing sheared axial flow. $\rho_0(r)$ is the mean flow density (here taken constant), and $U(r)$ is the mean flow velocity, here taken to be uniform outside a boundary layer of width $h$. $Z$ is the boundary impedance and defines the boundary condition at the wall of the duct.}
    \label{figCyl}
\end{figure}
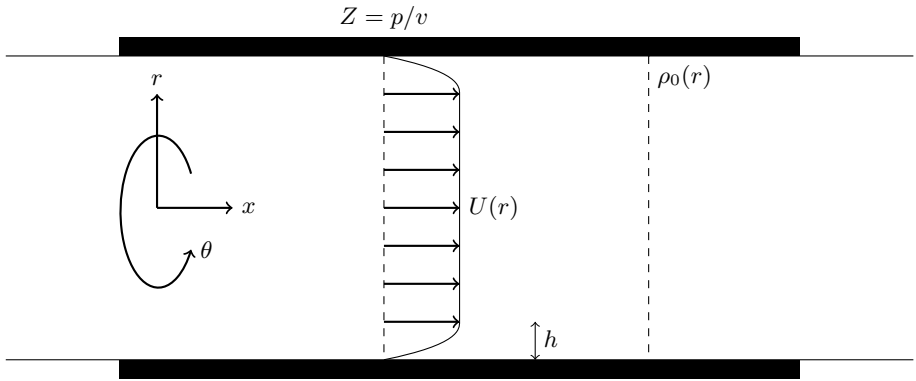

In what follows, we make the simplifying assumption of a constant density $\rho_0(r)$.  This is a homentropic assumption, and implies that $c_0(r)$ is also constant.  We may then nondimensionalize speeds by the sound speed $c_0$, densities by $\rho_0$, and distances by the duct radius $a$.  Note that this places the impedance boundary condition in nondimensional terms at $r=1$.  We also assume a flow profile $U(r)$ that is uniform, except within a boundary layer of width $h$ where it varies quadratically:
\begin{equation}\label{FlowProfile}
    U(r)=\begin{cases}M & 0\leq r\leq 1-h  \\
    M(1-(1-\frac{1-r}{h})^2) & 1-h\leq r \leq 1\end{cases}.
\end{equation}
With the nondimensionalization of velocities by $c_0$, $M$ here is the duct centreline Mach number.  This situation is depicted schematically in figure~\ref{figCyl}.

In order to solve the Pridmore-Brown equation~\eqref{PridmoreBrown}, we first consider solutions to the homogeneous form
\begin{equation}\label{HomogPridmoreBrown}
    \tp^{\prime\prime}+\lr{\frac{2k U^\prime}{\omega-U(r)k}+\frac{1}{r}}\tp^\prime+\lr{(\omega-U(r)k)^2-k^2-\frac{m^2}{r^2}}\tp=0.
\end{equation}

\subsection{Homogeneous Solutions Within the Region of Uniform Flow}
\label{Sect:Homog:Uniform}

Within the region of uniform flow, the homogeneous Pridmore-Brown equation~\eqref{HomogPridmoreBrown} reduces to
\begin{equation}\label{PridmoreBrownUnif}
    \tp^{\prime\prime}+\frac{1}{r}\tp^\prime+\lr{(\omega-Mk)^2-k^2-\frac{m^2}{r^2}}\tp=0.
\end{equation}
This is Bessel's equations of order $m$ rescaled by $\alpha$, where
\begin{equation}\label{alpha}
    \alpha^2=(\omega-Mk)^2-k^2;
\end{equation}
it will turn out later that the branch chosen for $\alpha$ does not matter, although for definiteness one may choose $\Real(\alpha)>0$.  Bessel's equation has two pairs of linearly independent solutions that we shall make use of: the Bessel functions of the First and Second kind, $J_m(\alpha r)$ and $Y_m(\alpha r)$; and the Hankel functions of the first and second kind. $H_m^{(1)}(\alpha r)$ and $H_m^{(2)}(\alpha r)$.  
More information regarding these can be found in \citet{abramowitz1965handbook}. It is worth noting that only $J_m(\alpha r)$ is regular at $r=0$, with the other solutions all requiring a branch cut along $\alpha r<0$, with a singularity at $\alpha r=0$.

\subsection{Homogeneous Solutions Within the Region of Sheared Flow}\label{Sect:Homog:Shear}
In this section, we will construct the solution to the homogeneous Pridmore-Brown equation~\eqref{HomogPridmoreBrown} when $U(r)$ varies by proposing a Frobenius expansion about the singularities of the Pridmore-Brown equation.

In addition to the singularity at $r=0$, the homogeneous Pridmore-Brown equation possesses regular singularities whenever $\omega - U(r)k = 0$; these singularities correspond to the critical layer.  Within the sheared flow region $1-h<r<1$, since the velocity profile $U(r)$ is quadratic in $r$, there are exactly two critical values $r = r_c$ for which $\omega-U(r_c)k=0$.  Note that in general these critical values will be complex.  Solving this quadratic equation gives the two singularities explicitly as $r_c^+$ and $r_c^-$, where
\begin{align}\label{rcplusminus}
    r_c^\pm&=1-h\pm Q &  Q&=h\sqrt{1-\frac{\omega}{Mk}}.
\end{align}
For convenience, we will take $\Real(Q)\geq0$, so that $\Real(r_c^+) \geq 1-h$ and $\Real(r_c^-) \leq 1-h$.  Since solutions with this quadratic flow profile $U(r)$ are only valid for $1-h<r<1$, it will therefore be $r_c^+$ that we are mostly concerned about here.

Following~\citet{brambley2012critical}, we propose a Frobenius expansion~\citep{teschl2012ordinary} about the regular singularity $r_c^+$,
\begin{align}\label{FrobeniusMethod}
    \tp(r)&=\sum_{n=0}^\infty a_n(r-r_c^+)^{n+\sigma} &
    &\text{with}\qquad a_0 \neq 0.
\end{align}

Specifying that $a_0 \neq 0$ results in a condition on $\sigma$, and we find that $\sigma=0,3$.  By Fuchs theorem~\citep{teschl2012ordinary}, this gives a pair of linearly independent solutions of the form
\begin{subequations}\label{ShearSolnCrit}\begin{align}
    \tp_1(r)&=\sum_{n=0}^\infty a_n (r-r_c^+)^{n+3}, \\
    \tp_2(r)&=A\tp_1(r)\log(r-r_c^+)+\sum_{n=0}^\infty b_n(r-r_c^+)^n.
\label{ShearSolnCritP2}
\end{align}\end{subequations}
The coefficients $a_n$ and $b_n$ are derived in appendix~\ref{appendix:frobenius}, where, in particular, it is found that
\begin{align}
    a_0 &= b_0 = 1, &
    b_1 &= 0, &
    b_2 = &-\frac{1}{2}\!\lr{\!k^2+\lr{\frac{m}{r_c^+}}^{\!\!2}}\!, &
    b_3 &=0,
\end{align}
and that
\begin{equation}\label{A}
    A=-\frac{1}{3}\lr{\frac{1}{Q}-\frac{1}{r_c^+}}\!\lr{\!k^2 + \lr{\frac{m}{r_c^{+}}}^{\!2}} - \frac{2m^2}{3r_c^{+3}},
    \end{equation}
the latter in agreement with equations~(2.3)--(2.5) of \citet{brambley2012critical}.  We note in passing that in practice we may be limited by the radius of convergence of~\eqref{ShearSolnCrit}, and in such cases the solutions given above are analytically continued by a companion expansion of the Pridmore-Brown equations about $r=1$, as described in appendix~\ref{appendix:frobenius1}.  Other than being a complication concerning numerical convergence, this complication may be ignored, and $\tp_1$ and $\tp_2$ thought of as being defined by the expressions in~\eqref{ShearSolnCrit}.

\begin{figure}
    \centering
\begin{tikzpicture}[scale=1.2]
\draw (0,5) rectangle (5,0);
\draw (1,1.25) --(1.25,1.25) node[anchor= north east] {$0$};
\draw[thick,->] (1,1.25) -- (4,1.25) node[anchor= west] {$\Real(r)$};
\draw[thick,->] (1.25,1) -- (1.25,4) node[anchor= south] {$\Imag(r)$};
\draw[thick] (2.5,1.25+0.1) -- (2.5,1.25-0.1) node[anchor= north] {$1-h$};
\draw[thick] (3.75,1.25+0.1) -- (3.75,1.25-0.1) node[anchor= north] {$1$};
\fill (2.5+0.5,1.25+0.25) circle (0.05cm) node[anchor= south west] {$r_c^+$};
\fill (2.5-0.5,1.25-0.25) circle (0.05cm) node[anchor= north east] {$r_c^-$};
\draw[black!60!green, thick, dashed] (2.5+0.5,1.25+0.25) -- (2.5+0.5,5);
\node[anchor = west, black!60!green] at (2.5+0.5,3) {Branch cut};
\draw[ultra thick] (2.5,1.25) -- (3.75,1.25);
\node[anchor = north east] at (4.75,4.75) {(a)};

\draw (0+6,5) rectangle (5+6,0);
\draw (1+6,1.25) --(1.25+6,1.25) node[anchor= north east] {$0$};
\draw[thick,->] (1+6,1.25) -- (4+6,1.25) node[anchor= west] {$\Real(r)$};
\draw[thick,->] (1.25+6,1) -- (1.25+6,4) node[anchor= south] {$\Imag(r)$};
\draw[thick] (2.5+6,1.25+0.1) -- (2.5+6,1.25-0.1) node[anchor= north] {$1-h$};
\draw[thick] (3.75+6,1.25+0.1) -- (3.75+6,1.25-0.1) node[anchor= north] {$1$};
\fill (2.5+0.5+6,1.25-0.25) circle (0.05cm) node[anchor= north west] {$r_c^+$};
\fill (2.5-0.5+6,1.25+0.25) circle (0.05cm) node[anchor= south east] {$r_c^-$};
\draw[black!60!green, thick, dashed] (2.5+0.5+6,1.25-0.25) -- (2.5+0.5+6,0);
\node[anchor = west, black!60!green] at (2.5+0.5+6,0.25) {Branch cut};
\draw[ultra thick] (2.5+6,1.25) -- (3.75+6,1.25);
\node at (4.75+6,4.75) {(b)};
\end{tikzpicture}
    \caption{Schematic of possible locations of the $r_c^+$ branch cut in the complex $r$-plane.  (a)~A possible choice of branch cut when $\Imag(r_c^+)>0$.  (b)~The other choice of branch cut is needed when $\Imag(r_c^+)<0$.}
    \label{figRad1}
\end{figure}
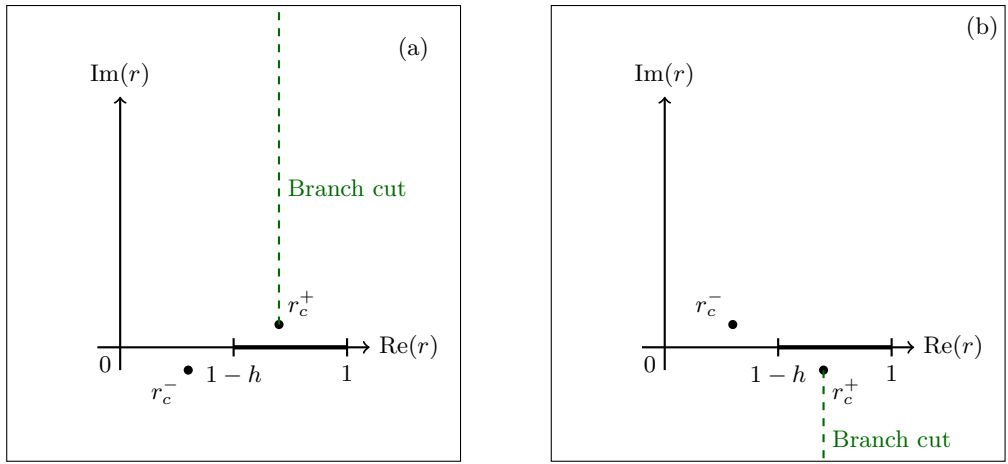

Due to the log term in $\tp_2$ in~\eqref{ShearSolnCritP2}, a branch cut is necessary in the complex $r$ plane originating from the branch point $r = r_c^+$. This branch cut must be such that the solutions remain continuous for the real values of $r\in[1-h,1]$, and so the branch cut must avoid crossing the real $r$ axis between $1-h$ and $1$. In the following, we achieve this by choosing the branch cuts parallel to the imaginary axis and away from the real axis, as depicted in figure~\ref{figRad1}.
When $r_c^+$ is real and $1-h<r_c^+<1$, no suitable choice of branch cut exists, and as a result any solution $\tp(r)$ with $\tp(r_c^+)\neq 0$ necessarily has a singular third derivative at $r_c^+$.  This only occurs for particular values of $k$, however, and we can map the corresponding values of $k$ in the complex $k$ plane to find they fall exactly on the half line $[\frac{\omega}{M},\infty)$; this range of excluded values of $k$ we refer to as the critical layer branch cut.
As $r_c^+$ becomes real, note that the value of $\tp_2(r_c^+)$ is different depending on whether we approach from positive or negative imaginary part.  Thinking of $r_c^+(k)$ as a function of $k$, this corresponds to approaching the critical layer branch cut $[\bk,\infty)$ in $k$ from above or below. This re-enforces the consideration of the critical layer appearing as a branch cut in the complex $k$ plane. The change in $\tp_2$ when crossing the critical layer branch cut from below to above is described as
\begin{equation}\label{JumpP2}
\Delta\tp_2(r)\;=\;\lim_{\mathclap{\mathrm{Im}(k)\searrow0+}}\;\tp_2(r)\;-\;\lim_{\mathclap{\mathrm{Im}(k)\nearrow0-}}\;\tp_2(r)\;=\; -2\pi \mi A \tp_1(r)H(r_c^+-r)
\end{equation}
Where $H(r)$ is the Heaviside function.

In order to retrieve this result we need only consider the $\log(r-r_c^+)$ term of $\tp_2$.  Note that $\partial r_c^+/\partial k > 0$ for real $k$ and real positive $\omega$; hence, if $k$ is nearly real and $\Imag(k)>0$, then $\Imag(r_c^+)>0$, and we must take the branch cut of $\log(r-r_c^+)$ upwards towards $+\I\infty$.  Similarly, for $\Imag(k)<0$ then $\Imag(r_c^+)<0$, and the branch cut for $\log(r-r_c^+)$ must be taken downwards to $-\I\infty$.  When $r_c^+$ is located on the real line, $(r-r_c^+)$ is negative for $r<r_c^+$. When we choose the branch cut into the upper half plane, this corresponds to a complex argument of $-\upi$.  When we choose the branch cut into the lower half plane, this corresponds to a complex argument of $\upi$.  This difference results in the jump of $2\upi\I$ given. If we instead consider $r>r_c^+$, the same argument is retrieved regardless of which direction we take the branch-cuts, and so no jump is observed.  This is the reason for the presence of the Heaviside function.

\subsection{Homogeneous Solutions Across the Full Domain}\label{SectFullDom}

In order to construct a full solution in $r\in[0,1]$, we now construct two solutions $\psi_1(r)$ and $\psi_2(r)$ that solve~\eqref{HomogPridmoreBrown} across $r\in[0,1]$, by patching together the solutions derived above in sections~\ref{Sect:Homog:Uniform} and~\ref{Sect:Homog:Shear}.  We construct $\psi_1(r)$ to satisfy the boundary condition at $r=0$~\eqref{r=0BndCon}, by taking
\begin{equation}\label{psi1}
\psi_1(r)=\begin{cases}
J_m(\alpha r) & 0\leq r\leq 1-h \\
C_1\tp_1(r)+D_1\tp_2(r) & 1-h\leq r\leq 1,
\end{cases}
\end{equation}
where the coefficients $C_1$ and $D_1$ ensure $C^1$ continuity, and are given by
\begin{subequations}\label{C1D1}\begin{align}
    C_1&=\phantom{-}\frac{J_m(\alpha(1-h))\tp_2^\prime(1-h)-\alpha J_m^\prime(\alpha(1-h))\tp_2(1-h)}{W(1-h)}, \\
    D_1&={-}\frac{J_m(\alpha(1-h))\tp_1^\prime(1-h)-\alpha J_m^\prime(\alpha(1-h))\tp_1(1-h)}{W(1-h)},
\end{align}\end{subequations}
and $W(r) = \mathcal{W}(\tp_1,\tp_2;r)$ is the Wronskian of $\tp_1$ and $\tp_2$, given in appendix~\ref{appendix:Sect:Wr} as
\begin{equation}\label{Wr}
    W(r)
    =\mathcal{W}(\tp_1,\tp_2;r)
    = \tp_1(r)\tp_2'(r)-\tp_{2}(r)\tp_1'(r)
    = -\frac{3}{4}\frac{r_c^+(r-r_c^+)^2(r-r_c^-)^2}{rQ^2}.
\end{equation}

Having constructed $\psi_1$ to satisfy the boundary condition at $r=0$, we now proceed to construct $\psi_2$ which satisfies the boundary condition~\eqref{r=1BndCon} at $r=1$.  Writing $\psi_2$ in terms of the homogeneous solutions derived above,
\begin{equation}\label{psi2}
    \psi_2(r)=\begin{cases}
    \cl_2H_m^{(1)}(\alpha r)+\dl_2H_m^{(2)}(\alpha r) & 0\leq r\leq 1-h \\
    \cg_2\tp_1(r)+\dg_2\tp_2(r) & 1-h\leq r \leq 1,
    \end{cases}
\end{equation}
we choose $\cg_2$ and $\dg_2$ to satisfy $\psi_2(1)=1$ and $\psi^\prime_2(1)=-\frac{\mi\omega}{Z}$.  This forces a non-zero normalized solution to $\psi_2$ which satisfies the boundary condition~\eqref{r=1BndCon} at $r=1$, and leads to
\begin{align}\label{C2gD2g}
\cg_2&= \frac{{\tp_2'(1)}+\frac{\mi\omega}{Z}{\tp_2(1)}}{W(1)}, &
\dg_2&= -\frac{{\tp_1'(1)}+\frac{\mi\omega}{Z}{\tp_1(1)}}{W(1)}.
\end{align}

The coefficients $\cl_2$ and $\dl_2$ are chosen such that our solution is $C^1$ continuous at $r=1-h$, giving
\begin{equation}\label{C2lD2l}
\!\lr{\!\!\begin{array}{c}
     \cl_2  \\
     \dl_2 
\end{array}\!\!\!}\!=\frac{\mi\upi(1\!-\!h)}{4}\!\lr{\!\!\!\begin{array}{cc}
    \phantom{-}\alpha H_m^{(2)\prime}\big(\alpha (1\!-\!h)\big) & \!\!\!{-}H_m^{(2)}\big(\alpha (1\!-\!h)\big)  \\
    {-}\alpha H_m^{(1)\prime}\big(\alpha (1\!-\!h)\big) & \!\!\!\phantom{-}H_m^{(1)}\big(\alpha (1\!-\!h)\big)
\end{array}\!\!\!}\!\!\lr{\! \begin{array}{cc}
    \tp_1(1\!-\!h) & \tp_2(1\!-\!h)  \\
    \tp_1^\prime(1\!-\!h) & \tp_2^\prime(1\!-\!h)
\end{array}\!\!}\! \lr{\!\!\begin{array}{c}
     \cg_2  \\
     \dg_2 
\end{array}\!\!\!}\!
\end{equation}
where the factor at the beginning comes from the Wronskian of $H_m^{(1)}$ and $H_m^{(2)}$ from \citet[][formula~9.1.17]{abramowitz1965handbook}.

We will also require later the jump in behaviour of $\psi_1$ and $\psi_2$ as $k$ crosses the critical layer branch cut from below to above.  Since any jump comes from the log term in $\tp_2(r)$ when $r<r_c^+$, we have, provided $r_c^+<1$,
\begin{subequations}\label{Jumps}\begin{gather}\label{JumpC1}
    \Delta C_1=2\mi\upi A D_1,
    \qquad\qquad
    \Delta\cg_2 = \Delta D_1 = \Delta\dg_2 = 0,
\\\label{jumpC2lD2l}
    \lr{\!\begin{array}{c}
     \Delta\cl_2  \\
     \Delta\dl_2 
\end{array}\!}=\frac{\upi^2(1\!-\!h)A \dg_2}{2}  \lr{ \begin{array}{cc}
    \phantom{-}\alpha H_m^{(2)\prime}\big(\alpha (1\!-\!h)\big) & {-}H_m^{(2)}\big(\alpha (1\!-\!h)\big)  \\
    {-}\alpha H_m^{(1)\prime}\big(\alpha (1\!-\!h)\big) & \phantom{-}H_m^{(1)}\big(\alpha (1\!-\!h)\big)
\end{array}}\lr{ \begin{array}{c}
\tp_1(1\!-\!h)  \\\tp_1^\prime(1\!-\!h)
\end{array}}
\end{gather}\end{subequations}
resulting in (provided $r_c^+<1$)
\begin{subequations}\label{jumptpsi1andpsi2}\begin{align}
    \Delta\psi_1(r)&=2\mi\upi AD_1\tp_1H(r-r_c^+), \\
    \Delta\psi_2(r)&=\begin{cases}
    \Delta\cl_2H_m^{(1)}(\alpha r)+\Delta\dl_2H_m^{(2)}(\alpha r) & 0\leq r\leq 1-h \\
    -2\mi\upi A \tp_1(r)\dg_2 H(r_c^+-r) & 1-h\leq r \leq 1.
    \end{cases}
\end{align}\end{subequations}
Note that, if $r_c^+>1$, then $\Delta\psi_1 = \Delta\psi_2 = 0$, since the $\psi_1$ and $\psi_2$ solutions are uniquely defined by their boundary conditions and no branch point occurs on the interval $r\in[1-h,1]$ to cause a jump.

\subsection{Modal solutions}
\label{ModalPoles}

Modal solutions of the homogeneous Pridmore-Brown equation~\eqref{HomogPridmoreBrown} are nonzero solutions $\tp(r)$ that satisfy both the boundary conditions at $r=0$ and at $r=1$~(\ref{r=0BndCon},\ref{r=1BndCon}).  In general, satisfying both boundary conditions would force the solution $\tp(r) \equiv 0$, so nonzero solutions exist only for particular modal eigenvalues $k$ (assuming $\omega$ is given and fixed).  In contrast, the solution $\psi_1(r)$ is never identically zero and always satisfies the homogeneous Pridmore-Brown equation and the boundary condition at $r=0$; indeed, any solution satisfying the boundary condition at $r=0$ is necessarily a multiple of $\psi_1(r)$.  Likewise, the solution $\psi_2(r)$ is never identically zero and always satisfies the homogeneous Pridmore-Brown equation and the boundary condition at $r=1$, and any solution satisfying the boundary condition at $r=1$ is necessarily a multiple of $\psi_2(r)$.  In general, $\psi_1$ and $\psi_2$ are linearly independent, and so their Wronskian $\mathcal{W}(\psi_1,\psi_2;r)$ is not identically zero, where
\begin{equation}\label{fullWron}
    \mathcal{W}(\psi_1,\psi_2;r)=\psi_1(r)\psi_2^\prime(r)-\psi_2(r)\psi_1^\prime(r).
\end{equation}
However, if $\tp(r)$ is nonzero and satisfies both boundary conditions at $r=0$ and $r=1$, then $\tp(r) = a\psi_1(r) = b\psi_2(r)$ for some nonzero coefficients $a, b$.  In other words, a modal solution is one where $\psi_1$ and $\psi_2$ are linearly dependent, and so $\mathcal{W}(\psi_1,\psi_2;r) \equiv 0$.

For $1-h\leq r\leq 1$, substituting $\psi_1$ from~\eqref{psi1} and $\psi_2$ from~\eqref{psi2} into the Wronskian~\eqref{fullWron} gives
\begin{equation}\label{WroninW}
    \mathcal{W}(\psi_1,\psi_2;r)=(C_1\dg_2-\cg_2D_1)W(r),
\end{equation}
where $W(r)$ is the Wronskian between $\tp_1$ and $\tp_2$ and is given earlier in~\eqref{Wr}. Since $\tp_1$ and $\tp_2$ were constructed to be linearly independent, we expect $W(r)$ not to be identically zero, and indeed~\eqref{Wr} shows that $W(r) \neq 0$ except at the critical layer $r = r_c^+$.  A modal solution, therefore, is given by the condition that $C_1\dg_2 - \cg_2 D_1 = 0$, which is independent of $r$, and implies that $C_1/D_1 = \cg_2/\dg_2$ and so that $\psi_1$ and $\psi_2$ are multiples of one another.

The same can be seen for $r\leq 1-h$.  In this case, the Wronskian~\eqref{fullWron} becomes
\begin{equation}\label{fullunif}
    \mathcal{W}(\psi_1,\psi_2;r)\;=\;\alpha\cl_2\mathcal{W}(J_m,H_m^{(1)};r)+\alpha\dl_2\mathcal{W}(J_m,H_m^{(2)};r) \;=\;\alpha(\cl_2-\dl_2)\frac{2\mi}{\pi r},
\end{equation}
where we have made use of the Bessel function identities~9.1.3, 9.1.4 and~9.1.16 from \citet{abramowitz1965handbook}.  Note in particular that $r \mathcal{W}(\psi_1,\psi_2;r)$ is a constant independent of $r$ for $0\leq r\leq 1-h$.  Since $\mathcal{W}(\psi_1,\psi_2;r)$ is continuous in $r$ across $r = 1-h$, since $\psi_1$ and $\psi_2$ are both $C^1$ continuous, it follows that for $0\leq r\leq 1-h$ we can set $r\mathcal{W}(\psi_1,\psi_2;r)=(1-h)\mathcal{W}(\psi_1,\psi_2;1-h)$.
We therefore arrive at the conclusion that
\begin{equation}\label{FullWronskian}
\mathcal{W}(\psi_1,\psi_2;r) = (C_1\dg_2-\cg_2D_1)\begin{cases}
W(r) & 1-h \leq r \leq 1 \\
W(1-h)\frac{1-h}{r} & 0 \leq r \leq 1-h,
\end{cases}\end{equation}
and that a mode corresponds to the dispersion relation $0 = D(k,\omega) = C_1\dg_2-\cg_2D_1$.  In the next section, we see how these modal solutions occur naturally as poles in the solution of the non-homogeneous Pridmore-Brown equation.

\section{Inhomogeneous Solutions and Inverting the Fourier Transform}
\label{section:greens}
\subsection{Inhomogeneous Solution to the Pridmore-Brown Equation}
While previously we have only been solving the homogeneous form \eqref{HomogPridmoreBrown}, our original problem was to solve the inhomogeneous Pridmore-Brown equation~\eqref{PridmoreBrown} subjected to a harmonic point mass source.  Due to the right hand side of \eqref{PridmoreBrown} being a scalar multiple of a delta function, located at $r=r_0$, our solution will be the same scalar multiple of the Green's function, and we denote this solution as $\tG$.  This function will satisfy the boundary condition at $r=0$ and $r=1$, and will solve the homogeneous Pridmore-Brown equation for $r<r_0$ and $r>r_0$; hence, $\tG$ may be written as a multiple of the homogeneous solution $\psi_1$ for $r<r_0$ and as a multiple of the homogeneous solution $\psi_2$ for $r>r_0$.  All that is required is to join the two solutions at $r=r_0$ such that they are continuous, and their derivative is discontinuous with a jump exactly matching the amplitude of the delta function.  This may be written succinctly as
\begin{equation}\label{tildeGfull}
    \tG=\frac{\omega-U(r_0)k}{2\upi\mi r_0}\frac{\psi_1(\rl)\psi_2(\rg)}{\mathcal{W}(\psi_1,\psi_2;r_0)},
\end{equation}
where
\begin{align}\label{rgrl}
    \rg&=\max(r,r_0), & \rl&=\min(r,r_0),
\end{align}
and once again $\mathcal{W}(\psi_1,\psi_2;r)$ is the Wronskian of $\psi_1$ and $\psi_2$.  Using~\eqref{FullWronskian}, this may be rewritten as
\begin{align}\label{tildeG}
    \tG&=\frac{\omega-U(r^*)k}{2\upi\mi r^*W(r^*)}\frac{\psi_1(\rl)\psi_2(\rg)}{C_1\dg_2-\cg_2D_1} &
    \text{where}\quad r^*&=\max(1-h,r_0)    
\end{align}

\subsection{Analytic continuation behind the critical layer branch cut}
\label{section:analytic-continuation}

The solution for $\tG$ in~\eqref{tildeG} above contains a branch cut along the critical layer $k\in[\bk,\infty)$.  We now introduce the following additional notation.  When evaluating a function $f(k)$ on the branch cut, for $k\in[\bk,\infty)$, we denote
\begin{align}
f^+(k) &= \lim_{\eps\to 0}f(k+\I\eps) &
f^-(k) &= \lim_{\eps\to 0}f(k-\I\eps) &
\Delta f(k) &= f^+(k) - f^-(k)
\end{align}
Note that the definition of $\Delta f$ agrees with the use of $\Delta$ in equations~(\ref{JumpP2},\ref{Jumps},\ref{jumptpsi1andpsi2}) above.  By using these equations, we find that
\begin{align}\label{JumptG}
\Delta\tG=&-\frac{\omega-U(r^*)k}{2\I\pi r^*W(r^*)}
\frac{1}{C^-_1\dg_2-\cg_2D_1+2\mi\upi AD_1\dg_2}\\
&\quad\times \left[\frac{2\mi\upi A D_1\dg_2\psi^-_1(\rl)\psi^-_2(\rg)}{C_1^-\dg_2-\cg_2D_1}-\psi^-_1(\rl)\Delta\psi_2(\rg)-\Delta\psi_1(\rl)\psi^-_2(\rg)-\Delta\psi_1(\rl)\Delta\psi_2(\rg)\right]\notag
.  
\end{align}

A typical branch cut, such as the branch cut in $\sqrt{z-z_0}$, may be taken in any direction from the branch point $z_0$.  The critical layer branch cut in the complex $k$-plane is different, in that the choice of branch cut was forced upon us by the requirement that the solution be continuous in $r$ for $r\in[1-h,1]$.  None-the-less, noting from~\eqref{JumptG} that $\Delta\tG$ is well defined function for general complex $k$, we may use equation~\eqref{JumptG} to analytically continue $\tG$ behind the critical layer branch cut.  For real $\omega$, we therefore define the analytic continuation of $\tG$ behind the branch cut into the lower-half $k$-plane as
\begin{equation}\label{tG+}
\tG^+(k) = \begin{cases}
\tG(k) & \Imag(k) > 0 \text { or } \Real(k) < \bk,\\
\tG(k)+\Delta G(k) & \Imag(k) < 0 \text { and } \Real(k) > \bk.
\end{cases}
\end{equation}

Similarly, we may rewrite~\eqref{JumptG} as
\begin{align}\label{JumptG+}
\Delta\tG=&-\frac{\omega-U(r^*)k}{2\I\pi r^*W(r^*)}
\frac{1}{C^+_1\dg_2-\cg_2D_1-2\mi\upi AD_1\dg_2}
\\&\quad\times
\left[\frac{2\mi\upi A D_1\dg_2\psi^+_1(\rl)\psi^+_2(\rg)}{C_1^+\dg_2-\cg_2D_1}-\psi^+_1(\rl)\Delta\psi_2(\rg)-\Delta\psi_1(\rl)\psi^+_2(\rg)+\Delta\psi_1(\rl)\Delta\psi_2(\rg)\right],\notag
\end{align}
which allows the analytic continuation of $\tG$ into the upper-half $k$-plane,
\begin{equation}\label{tG-}
\tG^-(k) = \begin{cases}
\tG(k) & \Imag(k) < 0 \text { or } \Real(k) < \bk,\\
\tG(k)-\Delta G(k) & \Imag(k) > 0 \text { and } \Real(k) > \bk.
\end{cases}
\end{equation}

The utility of these analytic continuations in not readily apparent.  However, their use allows for poles of $\tG$, corresponding to modal solutions to the homogeneous Pridmore-Brown equation, to be tracked behind the branch cut, and in particular a possible hydrodynamic instability mode will later be found to be hidden behind the critical layer branch cut in certain cases.  Their use also allows the deformation of integral contours behind the critical layer branch cut, as will be needed for the steepest descent contours needed for the large-$x$ asymptotic evaluation of the inverse Fourier transform.

In what follows $k^+$ and $k^-$ denote modal poles, see section \ref{ModalPoles}, of only $\tG^+$ or $\tG^-$ respectively.

\subsection{Inverting the Fourier Transform}\label{sect:Inhomg:Inversion}

Having formulated $\tG$ as the solution to the inhomogeneous Pridmore-Brown equation~$\eqref{PridmoreBrown}$, to recover the actual pressure perturbation $\hat{p}(x,r,\theta)$, we are required to invert the Fourier transform and sum the Fourier series.
For a fixed azimuthal mode $m$ we invert the Fourier transform using the formula
\begin{equation}\label{FourierInversion}
    G(x,r;r_0,m)=\frac{1}{2\upi}\int_{\mathcal{C}} \tG(r,r_0,k,m)\ex^{-\mi k x}dk.
\end{equation}
Note, however, that the critical layer branch cut is located along the real-$k$ axis $k\in[\bk,\infty)$.  We are therefore required to be careful in choosing a suitable inversion contour $\mathcal{C}$.

\subsubsection{Choosing an Inversion Contour}
In order to choose the correct Fourier inversion contour $\mathcal{C}$, we appeal to the Briggs--Bers criterion \citep{briggs1964electron, bers1983space}.  The Briggs--Bers criterion, summarized below, invokes the notion of causality; that the cause of the disturbance (the delta function forcing) should occur before the effect (the disturbance $\hat{p}$), which is otherwise lost when considering a time-harmonic forcing, as we do here.  A more in-depth description is available in many places in the literature~\citep[e.g.][appendix~A]{brambley2009fundamental}.

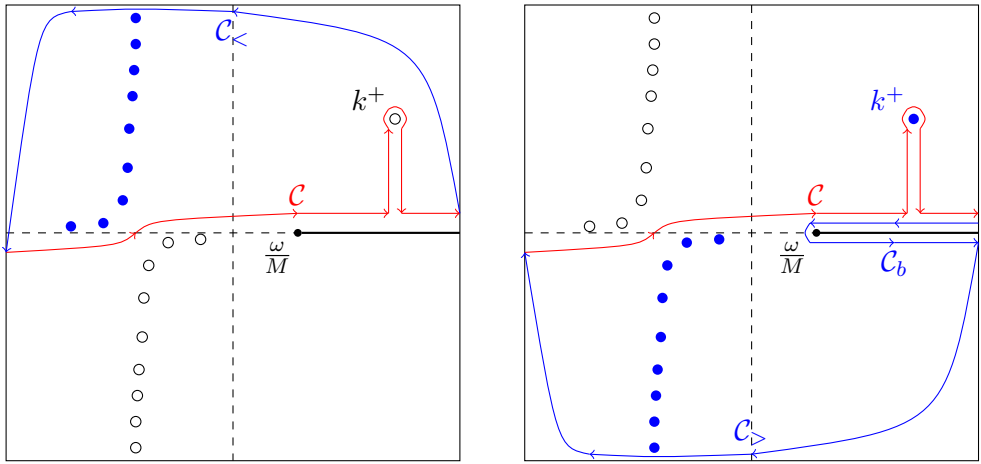
\begin{figure}
    \centering
\begin{tikzpicture} [scale=6/7]
\draw (-7.5,3.5) rectangle (-0.5,-3.5);
\draw (7.5,3.5) rectangle (0.5,-3.5);

\draw[dashed] (-7.5,0) -- (-3.06,0);
\draw[dashed] (0.5,0) -- (5,0);

\draw[dashed] (-4,3.5) -- (-4,-3.5);
\draw[dashed] (4,3.5) -- (4,-3.5);

\draw[thick] (-0.5,0) -- (-2.94,0) node[anchor= north east] {\large{$\bk$}};
\draw[thick] (7.5,0) -- (5,0) node[anchor= north east] {\large{$\bk$}};
\fill (-3,0) circle (0.06cm);
\fill (5,0) circle (0.06cm);

\fill[blue] (-6.5,0.1) circle (0.08cm);
\fill[blue] (-6,0.15) circle (0.08cm);
\draw (-4.5,-0.1) circle (0.08cm);
\draw (-5,-0.15) circle (0.08cm);

\fill[blue] (-5.7,0.5) circle (0.08cm);
\fill[blue] (-5.625,1) circle (0.08cm);
\fill[blue] (-5.6,1.6) circle (0.08cm);
\fill[blue] (-5.55,2.1) circle (0.08cm);
\fill[blue] (-5.525,2.5) circle (0.08cm);
\fill[blue] (-5.5,2.9) circle (0.08cm);
\fill[blue] (-5.5,3.3) circle (0.08cm);
\draw (-5.3,-0.5) circle (0.08cm);
\draw (-5.375,-1) circle (0.08cm);
\draw (-5.4,-1.6) circle (0.08cm);
\draw (-5.45,-2.1) circle (0.08cm);
\draw (-5.475,-2.5) circle (0.08cm);
\draw (-5.5,-2.9) circle (0.08cm);
\draw (-5.5,-3.3) circle (0.08cm);

\draw (-6.5,0.1)+(8,0) circle (0.08cm);
\draw (-6,0.15)+(8,0) circle (0.08cm);
\fill[blue] (-4.5,-0.1)+(8,0) circle (0.08cm);
\fill[blue] (-5,-0.15)+(8,0) circle (0.08cm);
\draw (-5.7,0.5)+(8,0) circle (0.08cm);
\draw (-5.625,1)+(8,0) circle (0.08cm);
\draw (-5.6,1.6)+(8,0) circle (0.08cm);
\draw (-5.55,2.1)+(8,0) circle (0.08cm);
\draw (-5.525,2.5)+(8,0) circle (0.08cm);
\draw (-5.5,2.9)+(8,0) circle (0.08cm);
\draw (-5.5,3.3)+(8,0) circle (0.08cm);
\fill[blue] (-5.3,-0.5)+(8,0) circle (0.08cm);
\fill[blue] (-5.375,-1)+(8,0) circle (0.08cm);
\fill[blue] (-5.4,-1.6)+(8,0) circle (0.08cm);
\fill[blue] (-5.45,-2.1)+(8,0) circle (0.08cm);
\fill[blue] (-5.475,-2.5)+(8,0) circle (0.08cm);
\fill[blue] (-5.5,-2.9)+(8,0) circle (0.08cm);
\fill[blue] (-5.5,-3.3)+(8,0) circle (0.08cm);

\draw (-1.5,1.75) circle (0.08cm) node[anchor= south east] {\large{$k^+$}};
\fill[blue] (6.5,1.75) circle (0.08cm) node[anchor= south east] {\large{$k^+$}};

\draw [red] (-1.6,1.6) .. controls  (-1.7,1.75) .. (-1.6,1.9);
\draw [red] (-1.6,1.9) .. controls (-1.5,1.95) .. (-1.4,1.9);
\draw [red] (-1.4,1.6) .. controls  (-1.3,1.75) .. (-1.4,1.9);
\draw [red, <-] (-1.6,1.6) -- (-1.6,0.3);
\draw [red, ->] (-1.4,1.6) -- (-1.4,0.3);
\draw [red, ->] (-1.4,0.3) -- (-0.5,0.3);
\draw [red, <-] (-1.6,0.3) -- (-3,0.3) node[anchor=south] {\large{$\mathcal{C}$}};
\draw [red, <-] (-3,0.3) .. controls  (-5.25,0.2) .. (-5.5,0);
\draw [red, ->] (-7.5,-0.3) .. controls  (-5.75,-0.2) .. (-5.5,0);

\draw [red] (6.4,1.6) .. controls  (6.3,1.75) .. (6.4,1.9);
\draw [red] (6.4,1.9) .. controls (6.5,1.95) .. (6.6,1.9);
\draw [red] (6.6,1.6) .. controls  (6.7,1.75) .. (6.6,1.9);
\draw [red, <-] (6.4,1.6) -- (6.4,0.3);
\draw [red, ->] (6.6,1.6) -- (6.6,0.3);
\draw [red, ->] (6.6,0.3) -- (7.5,0.3);
\draw [red, <-] (6.4,0.3) -- (5,0.3) node[anchor=south] {\large{$\mathcal{C}$}};
\draw [red, <-] (5,0.3) .. controls  (2.75,0.2) .. (2.5,0);
\draw [red, ->] (0.5,-0.3) .. controls  (2.25,-0.2) .. (2.5,0);

\draw [blue, ->] (-0.5,0.3) .. controls  (-1,3) .. (-4,3.4) node[anchor= north] {\large{$\mathcal{C_<}$}};
\draw [blue, ->] (-4,3.4) .. controls (-5.5,3.45) .. (-6.5,3.4);
\draw [blue, ->] (-6.5,3.4) .. controls  (-7,3.3) .. (-7.5,-0.3);

\draw [blue, <-] (4.9,0.15) -- (6.2,0.15);
\draw [blue, <-] (6.2,0.15) -- (7.5,0.15);
\draw [blue, ->] (4.9,-0.15) -- (6.2,-0.15) node[anchor=north] {\large{$\mathcal{C}_b$}};
\draw [blue, ->] (6.2,-0.15) -- (7.5,-0.15);
\draw [blue] (4.9,0.15) .. controls  (4.8,0) .. (4.9,-0.15);
\draw [blue, ->] (7.5,-0.15) .. controls  (7,-3) .. (4,-3.4) node[anchor= south] {\large{$\mathcal{C_>}$}};
\draw [blue, ->] (4,-3.4) .. controls (2.5,-3.45) .. (1.5,-3.4);
\draw [blue, ->] (1.5,-3.4) .. controls  (1,-3.3) .. (0.5,-0.3);

\end{tikzpicture}
    \caption{Illustration of typical pole locations, branch cuts, and inversion contours taken when an unstable $k^+$ pole is present for real $\omega$.  The inversion contour for $\tG$ is labelled $\mathcal{C}$. (Left) For $x<0$, the contour is closed in the upper half plane along the $\mathcal{C}_<$ contour.  (Right) For $x>0$, the contour is closed in the lower half plane along the $\mathcal{C}_>$ contour, and around the critical layer branch cut along the $\mathcal{C}_b$ contour.  Contributing modal poles are indicated in blue.}
    \label{fig:Contour}
\end{figure}

In order to make use to the Briggs--Bers criterion, the rate of exponential growth of the solution must be bounded; that is, there must exist $\Omega,K>0$ such that, if $\Imag(\omega)<-\Omega$, then $\tG$ is analytic for $|\Imag(k)|<K$.  For a given $\omega$ with $\Imag(\omega)<-\Omega$, we take the $k$-inversion contour $\mathcal{C}$ in~\eqref{FourierInversion} along the real-$k$ axis, and map the locations of any singularities (e.g.\ poles, branch points, etc). In order to find a correct integration contour for the real values of $\omega$ that are of interest, the imaginary part of $\omega$ is smoothly increased to $0$, and the locations of any singularities tracked throughout this process.  During this process, the $k$-inversion contour $\mathcal{C}$ must be smoothly deformed in order to maintain analyticity; that is, no singularities must cross the $k$-inversion contour.  Assuming this process may be completed and $\Imag(\omega)$ increased to zero, then the resulting $k$-inversion contour $\mathcal{C}$ is the correct causal contour.  Since for $x<0$ the $\exp\{-\I kx\}$ term is exponentially small as $|k| \to \infty$ in the upper-half $k$-plane, for $x<0$ we may close the contour with a large semi-circular arc at infinity in the upper-half $k$-plane, denoted $\mathcal{C}_>$.  The resulting contours, for real $\omega$, are illustrated in~\ref{fig:Contour},
for a typical unstable case.  The majority of singularities of $\tG$ are poles which do not cross the real $k$ axis as $\Imag(\omega)$ is varied, and hence correspond to exponentially decaying disturbances away from the point mass source at $x=0$.  The exception to these poles is the pole labelled $k^+$, which for this illustration originates in the lower-half $k$-plane for $\Imag(\omega)$ sufficiently negative, and therefore belongs below the $k$-inversion contour.  This implies that this pole is seen downstream of the point mass source, for $x>0$, despite having $\Imag(k)>0$, and therefore corresponds to an exponentially growing instability.  For a typical stable case, the situation is the same as shown in figure~\ref{fig:Contour} but with the $k^+$ pole not present.  Irrespective of the stability, the critical layer, as described earlier, exists when $k/\omega = 1/U(r_c) \in [1/M, \infty]$ for some critical radius $r_c$, and so is found in the lower-half $k$-plane for $\Imag(\omega)<0$.  Thus, as shown in figure~\ref{fig:Contour}, for $x>0$ in order to close $\mathcal{C}$ in the lower-half $k$-plane, we must pass around the critical layer branch cut, denoted by the contour $\mathcal{C}_b$, before closing in the lower-half $k$-plane with a semi-circular arc denoted $\mathcal{C}_>$.  The contribution from integrating around the critical layer branch cut, $\mathcal{C}_b$, leads to the non-modal contribution of the critical layer, and is discussed in detail below in section~\ref{Sect:Inhomog:Jumps}.

\subsubsection{Contribution from the poles of $\tG$}

We may now write the integral around the closed contour as a sum of residues of poles:
\begin{subequations}\label{IntegralContribs}\begin{align}
    \frac{1}{2\upi}\int_{\mathrlap{\mathcal{C}\cup\mathcal{C}_<}} \;\tG(r,r_0,k,m)\ex^{-\mi k x}dk &\;=\; G(x,r;r_0,m) \phantom{{}-I(x)} \;=\; \sum_{\mathclap{j:\;\Imag(k_j)>\Imag(\mathcal{C})}} R(k_j) &
    &\text{for} \quad x<0,
    \\
    \frac{1}{2\upi}\int_{\mathrlap{\mathcal{C}\cup\mathcal{C}_b\cup\mathcal{C}_>}} \;\tG(r,r_0,k,m)\ex^{-\mi k x}dk &\;=\; G(x,r;r_0,m) - I(x) \;=\; \sum_{\mathclap{j:\;\Imag(k_j) < \Imag(\mathcal{C})}} R(k_j) &
    &\text{for} \quad x>0,
\end{align}\end{subequations}
where $I(x)$ is the contribution from integrating around the critical layer branch cut contour $\mathcal{C}_b$ discussed in the next section, $R(k_j)$ is the residue from a pole at $k_j$ discussed below, and the notation $\Imag(k_j) > \Imag(\mathcal{C})$ is used to denote poles $k_j$ lying above the inversion contour $\mathcal{C}$.

The poles of $\tG$ correspond to zeros of the denominator of $\tG$, as given in~\eqref{tildeG}.  They can occur in two ways: as modal or non-modal poles. We consider the modal poles first.  The modal poles occur as zeros of the term $C_1\dg_2-\cg_2 D_1=0$.  As discussed in section~\ref{ModalPoles}, this occurs when both $\psi_1$ and $\psi_2$ satisfy both boundary conditions at $r=0$ and $r=1$.
These modal poles can be further classified into acoustic modes and surface modes: acoustic modes are those for which $\alpha$ in equation~\eqref{alpha} has a small imaginary part, and correspond to functions which are oscillatory in $r$; and surface modes are those for which $\alpha$ has a significant imaginary part, and correspond to functions which decay exponentially away from the duct walls at $r=1$.  For different parameters, we may find a variety of surface modes, and two with which we will be particularly interested here will be denoted $k^-$ and $k^+$.  For further details of surface modes, the reader is referred to the existing literature~\citep[e.g.][]{rienstra2003classification,brambley2013surface}.

Since the modal poles occur as zeros of $C_1\dg_2-\cg_2 D_1=0$, which we shall assume are simple zeros, the contribution from the residues of these poles are given by
\begin{equation}\label{ModalPoleContribution}
    R(k)=-\mathrm{sgn}(x)\frac{\omega-U(r^*)k}{2\upi r^* W(r^*)}\frac{\psi_1(\rl)\psi_2(\rg)}{\pfrac{}{k}\big(C_1\dg_2-\cg_2D_1\big)}\ex^{-\mi k x}.
\end{equation}

The second type of poles are the non-modal poles, which occur when $W(r^*)=0$. These occur when we loose independence between $\tp_1$ and $\tp_2$ at $r^*$. Note from the formula for $W(r)$, equation~\eqref{Wr}, that $W(r^*) = 0$ implies that $r^* = r_c^+$.
Since $1-h\leq r^*\leq 1$ this can only occur when $k$ is located on the critical layer branch cut.  In what follows, we will refer to this non-modal pole as $k_0$.  Note that $k_0$ is a function of the radial location of the point source $r_0$ (through $r^*$), which is unlike the modal poles for which $k_j$ is independent of the value of $r_0$; this is one reason this $k_0$ pole is referred to as a non-modal pole.  However, since our closed contour goes around the critical layer branch cut (along contour $\mathcal{C}_b$), this pole is always excluded from the sum of residues in~\eqref{IntegralContribs} above, and only occurs within the calculation of $I(x)$, which we consider next.

\subsubsection{Contribution from the critical layer Branch Cut}\label{Sect:Inhomog:Jumps}

\begin{figure}
    \centering
\begin{tikzpicture}[scale=5/7]
\draw (-7.5,3.5) rectangle (-0.5,-3.5);
\draw (7.5,3.5) rectangle (0.5,-3.5);

\draw[dotted] (-0.5,0) -- (-7.5,0) node[anchor=south, rotate=90] {$\Imag(k)$};
\draw[dotted] (7.5,0) -- (0.5,0) node[anchor=south, rotate=90] {$\Imag(k)$};

\fill (-6,0) circle (0.08cm) node[anchor=south] {$\bk$};
\fill (2,0) circle (0.08cm) node[anchor=south] {$\bk$};
\fill (-3,0) circle (0.08cm) node[anchor=south] {$k_<$};
\fill (5,0) circle (0.08cm) node[anchor=south] {$k_<$};
\fill (-1.5,0) circle (0.08cm) node[anchor=south] {$k_>$};
\fill (6.5,0) circle (0.08cm) node[anchor=south] {$k_>$};
\fill (-4.5,-2) circle (0.08cm) node[anchor=west] {$k^+$};
\fill (3.5,-2) circle (0.08cm) node[anchor=west] {$k^+$};
\fill (-5,-1) circle (0.08cm) node[anchor=west] {$k^-$};
\fill (3,-1) circle (0.08cm) node[anchor=west] {$k^-$};

\draw[thick] (-6,0) -- (-0.5,0);
\draw[thick] (2,0) -- (7.5,0);

\draw [red,<-](-0.5, 0.1) -- (-1.35,0.1);
\draw [red,->](-2.85, 0.1) -- (-1.65,0.1);
\draw [red,<-](-3.15,0.1) -- (-5.85,0.1);
\draw [red,->](-0.5, -0.1) -- (-1.35,-0.1);
\draw [red,<-](-2.85, -0.1) -- (-1.65,-0.1);
\draw [red,->](-3.15, -0.1) -- (-5.85,-0.1);

\draw [green,->](5.1, -0.15) -- (5.1,-3.5);
\draw [blue,<-](4.9, -0.15) -- (4.9,-3.5);
\draw [red,<-](1.9, -1.1) -- (1.9,-3.5);
\draw [red,<-](1.9, -0.15) -- (1.9,-0.9);
\draw [blue,->](2.1, -0.15) -- (2.1,-1.9);
\draw [blue,->,dashed](2.1,-1.9) -- (3.35,-1.9);
\draw [blue,<-,dashed](2.1,-2.1) -- (3.35,-2.1);
\draw [blue,->](2.1, -2.1) -- (2.1,-3.5);
\draw [purple,->](6.6, -0.15) -- (6.6,-3.5);
\draw [green,<-](6.4, -0.15) -- (6.4,-3.5);
\draw [red,<-,dashed](1.9,-0.9) -- (2.85,-0.9);
\draw [red,->,dashed](1.9,-1.1) -- (2.85,-1.1);

\draw [red] (-1.35,0.1) .. controls (-1.5,0.2) .. (-1.65,0.1);
\draw [red] (-3.15,0.1) .. controls (-3,0.2) .. (-2.85,0.1);
\draw [red] (-5.85,0.1) .. controls  (-6,0.2) .. (-6.15,0.1);
\draw [red] (-6.15,0.1) .. controls (-6.2,0) .. (-6.15,-0.1);
\draw [red] (-5.85,-0.1) .. controls  (-6,-0.2) .. (-6.15,-0.1);
\draw [red] (-1.35,-0.1) .. controls (-1.5,-0.2) .. (-1.65,-0.1);
\draw [red] (-3.15,-0.1) .. controls (-3,-0.2) .. (-2.85,-0.1);
\draw [green] (6.4,-0.15) .. controls  (6.3,0) .. (6.4,0.15);
\draw [purple] (6.4,0.15) .. controls (6.5,0.2) .. (6.6,0.15);
\draw [purple] (6.6,-0.15) .. controls  (6.7,0) .. (6.6,0.15);
\draw [blue] (4.9,-0.15) .. controls  (4.8,0) .. (4.9,0.15);
\draw [green] (4.9,0.15) .. controls (5,0.2) .. (5.1,0.15);
\draw [green] (5.1,-0.15) .. controls  (5.2,0) .. (5.1,0.15);
\draw [blue,dashed] (3.35,-1.9) .. controls  (3.5,-1.8) .. (3.65,-1.9);
\draw [blue,dashed] (3.65,-1.9) .. controls (3.7,-2) .. (3.65,-2.1);
\draw [blue,dashed] (3.35,-2.1) .. controls  (3.5,-2.2) .. (3.65,-2.1);
\draw [red] (1.9,-0.15) .. controls  (1.8,0) .. (1.9,0.15);
\draw [blue] (1.9,0.15) .. controls (2,0.2) .. (2.1,0.15);
\draw [blue] (2.1,-0.15) .. controls  (2.2,0) .. (2.1,0.15);
\draw [red,dashed] (2.85,-0.9) .. controls  (3,-0.8) .. (3.15,-0.9);
\draw [red,dashed] (3.15,-0.9) .. controls (3.2,-1) .. (3.15,-1.1);
\draw [red,dashed] (2.85,-1.1) .. controls  (3,-1.2) .. (3.15,-1.1);
\end{tikzpicture}
    \caption{(Left) Illustration of the integration contour required for the computation of the contribution from the critical layer branch cut, understood by integrating above and below the branch cut. Possible poles of $\tG^-$ and $\tG^+$ are denoted $k^-$ and $k^+$ respectively.
    (Right) The integration contour after being transformed onto the steepest descent contour.  Red lines behave as if evaluated below $\bk$ (using $\tG^-$); blue as if having been analytically continued around the $\bk$ branch point; green as if having been analytically continued around the $\bk$ and $k_<$ branch points; and purple as if analytically continued around all branch points, giving $\tG^+$. Note that we have been required to deform contours around the $k^+$ and $k^-$ poles.}
    \label{figSteepest}
\end{figure}
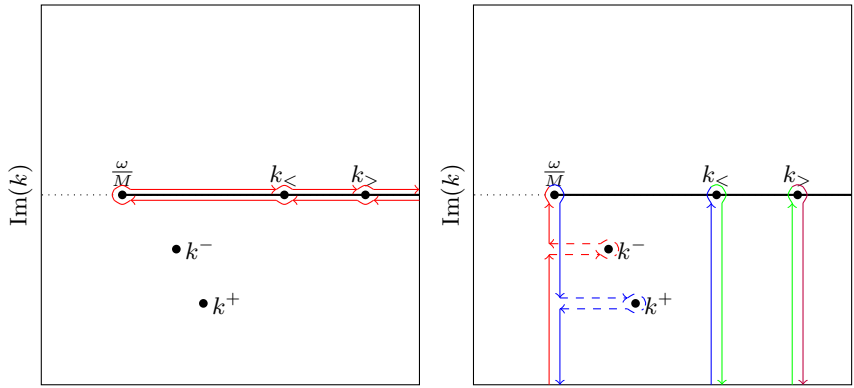

The contribution from the critical layer branch cut, including any non-modal pole $k_0$ along the branch cut, is contained solely within the integral along around the critical layer branch cut denoted $\mathcal{C}_b$ in figure~\ref{fig:Contour},
\begin{equation}
I(x) = \frac{-1}{2\upi}\int_{\mathcal{C}_b} \tG \ex^{-\mi k x}\,\intd k.
\end{equation}
However, as it stands, this integral for $I(x)$ is oscillatory, owing to the $\ex^{-\I kx}$ factor in the integrand, and so is difficult to accurately compute numerically.  This is especially true for large values of $x$.  Instead, it is helpful to deform the integral onto the Steepest Descent contour, for which $\ex^{-\I kx}$ is exponentially decaying along the contour.  This contour deformation has three benefits: firstly, it allows accurate numerical calculation of the integral; secondly, it allows the derivation of large-$x$ asymptotics using the Method of Steepest Descents; and thirdly, it brings insight into the various contributions that make up $I(x)$.  In deforming the integration contour, however, we must analytically continue $\tG$ behind the branch cut (as described in section~\ref{section:analytic-continuation} above), and carefully deform around any poles and branch points. This is illustrated schematically in figure~\ref{figSteepest}.
Note that poles and branch points of $\tG$ may exist behind the critical layer branch cut, and we must therefore use analytic continuations of $\tG$; the reader is reminded that $\tG^+$ is the analytic continuation of $\tG$ down behind the branch cut from above, while $\tG^-$ is the analytic continuation of $\tG$ up behind the branch cut from below.  Here, we use the notation that a pole of $\tG^+$ with $\Real(k)>\bk$ is denoted $k^+$, and a pole of $\tG^-$ with $\Real(k)>\bk$ is denoted $k^-$.  Thus, a $k^+$ pole with $\Imag(k^+)<0$ or a $k^-$ pole with $\Imag(k^-)>0$ are considered as being hidden behind the critical layer branch cut.  In the schematic in figure~\ref{figSteepest}, one $k^-$ and one $k^+$ pole are present, both with $\Imag(k)<0$, although this is not always the case; moreover, if present, the $k^+$ pole may interact with the integral contours around $k_<$ and $k_>$, in addition to interacting with the integral contour around $\bk$ depicted in figure~\ref{figSteepest}, depending on the location of the $k^+$ pole.

The Steepest Descent contours are where $\ex^{-\I kx}$ is exponentially decaying; i.e.\ towards $-\I\infty$ in the complex $k$ plane.  There is no difficulty deforming the contour at infinity, since $\ex^{-\I kx}$ is exponentially small there (provided $x>0$, which is the only case in which the critical layer branch cut contributes).  Along the branch cut there are up to three branch points singularities, denoted $\bk$, $k_<$ and $k_>$ in figure~\ref{figSteepest}, that must be deformed around.  These occur because of the presence of the $\log(r-r_c^+)$ term in $\tp_2(r)$, and the presence of $\tp_2(1\!-\!h)$, $\tp_2(r_0)$ and $\tp_2(r)$ in the expression for $\tG$;  each of these terms leads to a branch point, respectively at $\bk$, at $k_0$ corresponding to $r_c^+(k_0) = r_0$, and at $k_r$ corresponding to $r_c^+(k_r) = r$. 

Moreover, $\tG$ possesses a pole at $k_0$, which is exactly the non-modal pole referred to above, although there are no poles of $\tG$ at $\bk$ or at $k_r$.  Details of these calculations are given in appendix~\ref{appendix:decay-rates}.  The branch point at $k_r$ is not present when $r \leq 1-h$, and the pole and branch point at $k_0$ are not present when $r_0 \leq 1-h$.  For simplicity in what follows, we denote $k_< = \min\{k_0, k_r\}$ and $k_> = \max\{k_0, k_r\}$, as depicted in figure~\ref{figSteepest}.

The total integral around the branch cut can therefore be found by summing these three integrals, subtracting any $k^-$ contributions below the branch cut and adding any $k^+$ contributions below the branch cut, and adding the pole residue at $k_0$ calculated as if it was located above the branch cut.  This results in
\begin{equation}\label{IntegralinSteepest}
    I(x)=I_{\bk}(x)+I_0(x)+I_r(x)+R_0^+(k_0)+\sum_{\mathclap{\Imag(k^+)<0}}R^+(k^+)-\sum_{\mathclap{\Imag(k^-)<0}}R^-(k^-),
\end{equation}
where $R^\pm$ is the residue given in~\eqref{ModalPoleContribution} evaluated using $\tG^\pm$, $R_0^+(k_0)$ is the residue of the non-modal pole $k_0$ evaluated as if approached from above the branch cut, derived in appendix~\ref{appendix:decay-rates-r0} and given in~\eqref{appendix:equ:R+} as
\begin{equation}
R_0^+(k_0) = \frac{2Mk_0^2(\omega-Mk_0)\ex^{-\mi k_0x}}{3\pi r_0h^2\omega(C_1^+\dg_2-\cg_2D_1)}
\begin{cases}
\dg_2\psi_1(r) & r < r_0\\
D_1\psi_2(r) & r > r_0,
\end{cases}
\end{equation}
the steepest descent integrals are defined as
\begin{equation}\label{Integralform}
    I_q(x)=\frac{1}{2\upi\mi}\int_0^\infty \Delta\tG_q(k_q-\mi\xi)\ex^{-\mi(k_q-\mi\xi)x}\,\intd\xi,
\end{equation}
and the jumps across each of the Steepest Descent branch cuts are calculated in appendix~\ref{appendix:DeltaGq} to be
\begin{subequations}\label{JumpBC}\begin{align}\label{JumptGbk}
\Delta\tG_{\bk}&=\frac{-\lr{\omega-U(r^*)k} A }{ r^*W(r^*) \big(C_1^-\dg_2-\cg_2D_1+2\mi\upi AD_1\dg_2\big)}\times\begin{cases}
\dfrac{\dg_2^2\psi_1^-(\rl)\psi_1^-(\rg)}{\lr{C_1^-\dg_2-\cg_2D_1}} &\rg<1\!-\!h \\\\
\dfrac{D_1\dg_2\psi_1^-(\rl)\psi_2^-(\rg)}{\lr{C_1^-\dg_2-\cg_2D_1}} & \rl<1\!-\!h<\rg \\\\
\dfrac{D_1^2\psi_2^-(\rl)\psi_2^-(\rg)}{\lr{C_1^-\dg_2-\cg_2D_1}} & 1\!-\!h<\rl,
\end{cases}
\\
\label{JumptGkless}
\Delta\tG_<&=\frac{-(\omega-U(r^*)k)}{ r^* W(r^*)}\frac{ A D_1 \tp_1(\rl)\psi_2^-(\rg) }{C_1^-\dg_2-\cg_2D_1+2\mi\upi A D_1\dg_2}H\big(\rl-(1-h)\big),
\\\notag\\
\label{JumptGkgreat}
\Delta\tG_>&=\frac{-(\omega-U(r^*)k)}{ r^* W(r^*)}\frac{ A \dg_2 \psi_1^-(\rl)\tp_1(\rg) }{C_1^-\dg_2-\cg_2D_1+2\mi\upi A D_1\dg_2}H\big(\rg-(1-h)\big),
\end{align}\end{subequations}

Note that $\Delta\tG_{\bk} + \Delta\tG_< + \Delta\tG_> = \Delta\tG = \tG^+ - \tG^-$.
While these integrals are now amenable to numerical integration, additional understanding of the contribution from the three steepest descent contours may be gained by considering the large-$x$ limit.

\subsection{Far-Field Decay Rates of the Critical Layer Contribution}\label{sect:Inhomg:Steepest}

The critical layer branch cut contribution~\eqref{IntegralinSteepest} contains integrals $I_q(x)$ given by~\eqref{Integralform} which are amenable to asymptotic analysis in the limit $x \to \infty$, using the Method of Steepest Descent.  Having already deformed the integration contours onto the steepest descent contours, so that the integrands have had their $x$-dependent oscillation removed and are now exponentially decaying along the contour, we may directly apply Watson's Lemma \citep{watson1918harmonic}.  If some function $q(k)$ satisfies $f(k_q-\mi\xi)\sim B\xi^\nu+O(\xi^{\nu+1})$ to leading order for small $\xi$ with $\nu>-1$, then Watson's Lemma implies that, for large $x$,
\begin{equation}\label{Watsons}
    \frac{1}{2\mi\upi}\int_0^\infty f(k_q-\mi\xi)\ex^{-\mi(k_q-\mi\xi)x}\,\intd\xi\sim\frac{B\Gamma(\nu+1)\ex^{-\mi k_q x}}{2\mi\upi x^{\nu+1}}+O\big(x^{-(\nu+2)}\big),
\end{equation}
where $\Gamma$ is the Gamma function, and in particular, $\Gamma(\nu+1) = \nu!$ for integer $\nu$.
For each of the $I_q(x)$ integrals, this can then be interpreted as an algebraically decaying wave of phase velocity $\frac{\omega}{k_q}$.

In order to find the decay rates of the steepest descent contours we are required to understand the behaviour of $\psi_1(r,k_q-\mi\xi)$ and $\psi_2(r,k_q-\mi\xi)$ for small $\xi$ at $r \in\{ 1-h,r,r_0,1\}$. Details of these can be found in appendix~\ref{appendix:decay-rates}.  The result, given in equations~\eqref{equ:appendix:tGbk} and~\eqref{equ:appendix:j}, is that, for $k=\bk-\mi\xi$, as $\xi\to 0$ with $\xi>0$, we find that
\begin{equation}\label{leadDtGbk}
    \Delta\tG_{\bk}\sim\begin{cases}
     \xi^{3/2} & r_0\leq 1-h \\
     \xi^{5/3} & r_0>1-h.
    \end{cases}
\end{equation}
By Watson's Lemma, this results in a wave convected with the flow speed $M = U(1-h)$ and algebraically decaying like $x^{-\frac{5}{2}}$ when the source is within the region of uniform flow, and $x^{-\frac{7}{2}}$ for a source located in the sheared flow; the pre-factor in each case is different, and is also governed by the above expressions.

In the case $r_0>1-h$, the leading order contribution to $\Delta\tG_0$ as $k\to k_0$ is derived in appendix~\ref{appendix:decay-rates-r0} as
\begin{equation}\label{DeltatG0lim}
\Delta\tG_0 = \frac{A\omega h^2U(r_0)}{6r_0Mk_0^2(r_0-1+h)}\frac{ (k-k_0)^2}{C_1^-\dg_2-\cg_2D_1+2\upi\mi AD_1\dg_2}
\times\begin{cases}
\dg_2\psi_1^-(r) & r_0 > r\\
D_1\psi_2^-(r) & r_0 < r.
\end{cases}
\end{equation}
By Watson's Lemma, this results in a wave convected with the flow speed at the point source, $U(r_0)$, and decaying algebraically like $x^{-3}$.

Finally, considering $\Delta\tG_r$ as $k\to k_r$, it is found in appendix~\ref{appendix:decay-rates-r} that
\begin{equation}\label{DeltatGrlim}
\Delta\tG_r\sim\frac{A(\omega-U(r^*)k_r)\omega^3 h^6}{8r^* W(r^*)M^3k_r^6(r-1+h)^3 }\frac{(k-k_r)^3}{C_1^-\dg_2-\cg_2D_1+2\mi\upi A D_1\dg_2}\times \begin{cases}
\dg_2 \psi_1^-(r_0) &  r_0 < r\\
D_1 \psi_2^-(r_0) & r_0 > r.
\end{cases}
\end{equation}
By Watson's Lemma, this results in a wave convected with the flow speed $U(r)$ and decaying algebraically like $x^{-4}$.

It may be noted that the decay rates for $I_0$ and $I_r$ are the same as predicted for a linear boundary layer flow profile by \citet{brambley2012critical}.  We now proceed to compare these results with previous literature.

\subsubsection{Comparisons with Previous Far-Field Scalings}

Our results for the large-$x$ decay of the various components of the critical layer are compared to those predicted by \citet{swinbanks1975sound} for a general flow profile, and those predicted by \citet{brambley2012critical} for a constant-then-linear flow profile, in table~\ref{tab:results}.
\begin{table}
\centering
\begin{tabular}{l|c|c|c|c|c}
                             & \multicolumn{2}{|c|}{$I_{\bk}$}           & $I_r$    & $I_{0}$   & $R_0^+(k_0)$ \\[0.5ex]
                             & $r_0\leq 1-h$      & $r_0>1-h$          & $r>1-h$  & $r_0>1-h$   & $r_0>1-h$\\\hline
\citeauthor{swinbanks1975sound}   & \---                 &   \---                 &   \---       & $x^{-3}$  & 1\\
Linear BL & $x^{-4}$           & $x^{-5}$           & $x^{-4}$ & $x^{-3}$  & 1\\
Quadratic BL                             & $x^{-\frac{5}{2}}$ & $x^{-\frac{7}{2}}$ & $x^{-4}$ & $x^{-3}$ & 1
\end{tabular}
\caption{Comparison of the different decay rates given for a general flow profile by \citet{swinbanks1975sound} and for a linear boundary layer flow profile by \citet{brambley2012critical} against those given here for a quadratic boundary layer flow profile.}
\label{tab:results}
\end{table}%
The $I_0$ integral gives a wave with phase velocity equal to that of the mean flow at the location of the point mass source, $U(r_0)$, provided the point mass source is in a region of sheared flow, $r_0>1-h$.  It can be observed in table~\ref{tab:results} that agreement is seen in all three works for $r_0>1-h$. While \citeauthor{swinbanks1975sound} did not consider the other cases in detail, this work finds further agreement for the $I_r$ contribution with \citeauthor{brambley2012critical}. In both the linear and quadratic shear flow cases, when the source is located within the region of sheared flow, the $I_0$ contribution is the slowest decaying term.  When the source is located within the uniform flow region, the $I_{\bk}$ contribution is the slowest decaying term, although this is matched by $I_r$ contribution for linear shear. It should noted, however, that when $r_0>1-h$ we have in addition the contribution of the non-modal $k_0$ pole, which does not decay.

The difference in the behaviour of the $I_{\bk}$ integrals may be understood as having two causes.  The first is the difference in behaviour of the constant $A$, given in general as
\begin{equation}
    A=-\frac{1}{3}\lr{\frac{\omega^2}{M^2}+\frac{m^2}{r_c^2}}\lr{\frac{U^{\prime\prime}(r_c)}{U^\prime(r_c)}-\frac{1}{r_c}}-\frac{2m^2}{3r_c^{3}}.
\end{equation}
In the case of linear shear flow, the $U^{\prime\prime}$ term is zero for $k\not=\bk$ and the resulting expression is $O(1)$ as $k\to\bk$. In the case of a quadratic shear boundary layer, $U^{\prime\prime}$ is non zero and dominates $A$ as $k\to\bk$, providing a factor of $(k-\bk)^{-\frac{1}{2}}$. The remainder of the differences between the decay rates is explained, for $I_{\bk}$, by the fact that $(r_c^+-(1-h))\sim(k-\bk)^{\frac{1}{2}}$ in the quadratic shear case, where as for linear shear $(r_c-(1-h))\sim(k-\bk)$.  For the $I_0$ and $I_r$ contributions, where we do not have $r_c^+\to 1-h$, and all the other terms are equivalent between the linear and quadratic cases, therefore giving the same eventual asymptotics scalings, although the pre-factors may vary significantly.  Further details are given in appendix~\ref{appendix:n-polynomial}.

\section{Numerical Results}
\label{section:results}

In this section, the above analysis is illustrated with some numerical examples.  The Frobenius series solutions~$\tp_1$ and~$\tp_2$ are computed by summing the terms of the series, as given in appendix~\ref{appendix:frobenius}, until a relative error of order $10^{-16}$ is achieved, using the Matlab code in the supplementary material.  In particular, this is more numerically expensive and prone to numerical rounding errors near the edge of the radius of convergence for each of the solutions, and care must therefore be taken to choose an accurate and  efficient numerical implementation  of $\tp_1$ and $\tp_2$ in terms of the Frobenius series solutions. The modal poles are found using a variant of the Secant method, and have been confirmed against results using a finite-difference method applied to the Pridmore-Brown equation~\citep{brambley2011well}.  When summing the residues of modal poles, all poles with $|\Imag(k)|<400$ have been included.

Throughout this section we show results from four parameter sets, detailed in table~\ref{tab:Params}.
\begin{table}
\centering
\begin{tabular}{r@{$\qquad$}c@{$\qquad$}c@{$\qquad$}c@{$\qquad$}c@{$\qquad$}c}
                                &           & \setA  & \setB  & \setD         & \setE     \\[5pt]
 Frequency                      & $\omega$  & 10     & 10     & 31            & 16        \\
 Azimuthal order                & $m$       & 5      &  5     & 24            & 24        \\
 Centreline Mach number         & $M$       & 0.5    & 0.5    & 0.5           & 0.35      \\
 Boundary layer thickness       & $h$       & 0.05   & 0.001  & 0.01          & 0.005     \\
 Impedance mass                 & $\mu$     & 0.2    & 0.2    & 0.01          & 0.01      \\
 Impedance spring               & $K$       & 10     & 10     & 10            & 10        \\
 Impedance damper               & $R$       & 2      &  2     & 0.75          & 0.75      \\
 Impedance $Z(\omega)$          & $Z$       & $2+\I$ & $2+\I$ & $0.75-0.01\I$ & $0.75-0.465\I$ 
\end{tabular}
\caption{Parameter sets used for the following numerical results.  The impedance is of mass--spring--damper type, $Z(\omega)=R+\mi \mu \omega -\mi K/\omega$.}
\label{tab:Params}
\end{table}%
These parameter sets are inspired by values used in previous studies~\citep{brambley2012critical,brambley2013surface,brambley2016time}, and motivated by application to aeroengine intakes; in particular, parameter set~\setD\ is intended to be typical of a rotor-alone tone at takeoff, while parameter set~\setE\ might represent the same type of mode during the landing approach.

In section~\ref{Sect:Nums:Poles}, we will explore the locations of the modal poles in the complex $k$-plane.  This section will particularly focus on the $k^\pm$ modal poles discussed in~\ref{sect:Inhomg:Inversion}, including tracking these modal poles as they move behind the critical layer branch cut for certain parameters (by taking advantage of the ability of the Frobenius series solutions to analytically continue behind the critical layer branch cut). In section~\ref{Sect:Nums:BC} we compare the various contributions from the critical layer branch cut described in section~\ref{Sect:Inhomog:Jumps}, including their large $x$ behaviour, and show that these agree with the predicted large $x$ behaviour from section~\ref{sect:Inhomg:Steepest}. In section~\ref{Sect:Nums:FullInv} the full solution in terms of $x$ and $r$ is plotted, and these results are compared to the linear boundary layer case. Finally, in section~\ref{Sect:Nums:Stab}, we investigate how the results vary as we vary parameters, including the frequency $\omega$, the boundary layer thickness $h$, the wall impedance $Z$, and the steady mean flow velocity $M$.

\subsection{Pole Locations.}
\label{Sect:Nums:Poles}

\begin{figure}
    \centering
\includegraphics{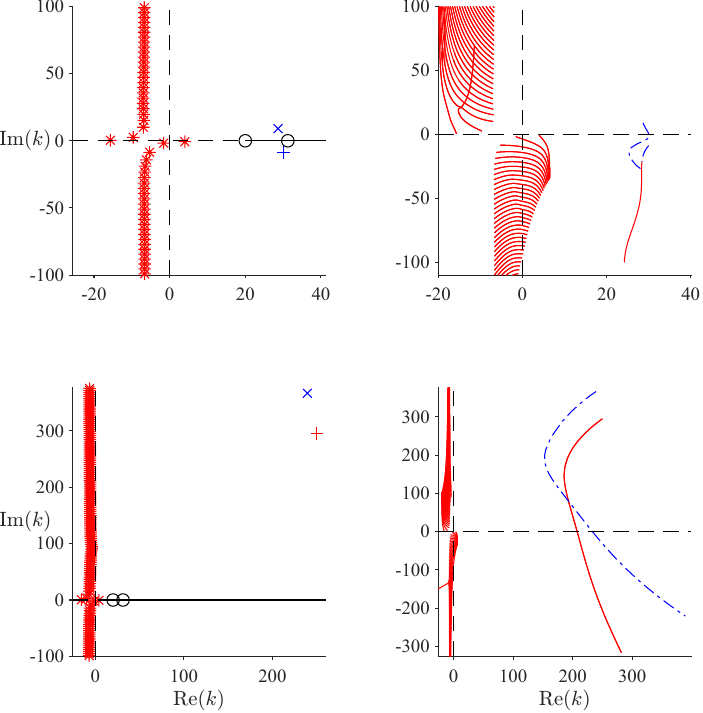}
    \caption{Location of the poles in the complex $k$ plane for parameter sets~\setA\ (top) and~\setB\ (bottom).  Left: For real $\omega$: acoustic modes with $\Real(k)<\bk$ $(\ast)$; $k^+$ poles $(+)$; $k^-$ poles $(\times)$; the critical layer branch cut (\---); and branch points $\bk$ and $k_0$ for $r_0=1-\frac{9h}{10}$ $(\circ)$. Right: Trajectories of poles for $-50 < \Imag(\omega) < 0$. Poles coloured red (left) and solid lines (right) denote poles contributing to the modal sum.  Poles coloured blue (left) and dashed lines (right) denote poles hidden behind the branch cut (which varies with $\Imag(\omega)$) and do not contribute.}
    \label{fig:PoleMovement}
\end{figure}

The locations of the poles in the complex $k$ plane for parameter sets~\setA and~\setB are plotted in figure~\ref{fig:PoleMovement}.
In addition to the usual acoustic modes (denoted as $\ast$ in figure~\ref{fig:PoleMovement}), one $k^+$ and one $k^-$ pole is found for each parameter set.  For parameter set~\setA, both the $k^+$ and $k^-$ poles are behind the critical layer branch cut, and so would not be found using conventional numerical methods, although the $k^+$ pole does still contribute to the total pressure field through the critical layer branch cut contribution, as described in section~\ref{Sect:Inhomog:Jumps} above.  In contrast, for parameter set~\setB, the $k^+$ pole is not behind the branch cut and takes the form of a standard modal pole, in this case a hydrodynamic instability surface wave.  The stability of the modal poles is verified from the movement of the poles in the $k$ plane as $\Imag(\omega)$ is decreased from zero, following the Briggs--Bers Criterion (shown in the right-hand plots in figure~\ref{fig:PoleMovement}); note that the critical layer branch cut also moves as a function of $\Imag(\omega)$.  Of particular note is that the $k^+$ pole for parameter set~\setA\ emerges from behind the critical layer branch cut as $\Imag(\omega)$ is reduced from zero, becoming a standard modal pole provided $\Imag(\omega)$ is taken sufficiently negative.

As discussed in section~\ref{sect:Inhomg:Inversion}, when the $k^+$ pole is located above the branch cut it is unstable, with a contribution growing exponentially in $x$.  When the $k^+$ pole is located below the branch cut we do not see its contribution to the modal sum directly, but instead it contributes as part of the branch cut integral, as seen when deforming onto the steepest descent contour. In this latter case, we would observe a contribution that decays in $x$. In both examples, the $k^-$ pole in located above the branch cut and does not contribute towards the Fourier inversion. In the event that this $k^-$ pole were located below the branch cut, its contribution would almost exactly cancel the critical layer branch cut contribution, again seen by deforming onto the steepest descent contour; however, the $k^-$ pole has not been found below the branch cut for any parameters considered here, unlike in the linear boundary layer profile case, which is investigated further in section~\ref{Sect:Nums:FullInv} below.

Also plotted in figure~\ref{fig:PoleMovement} is the critical layer branch cut for $k > \bk$, together with the non-modal $k_0$ pole, which is only present for a point mass source within the boundary layer, $r_0>1-h$.  The effects of these non-modal contributions are illustrated in the next section.

\subsection{Branch cut contributions}\label{Sect:Nums:BC}

\begin{figure}
\centering%
\includegraphics{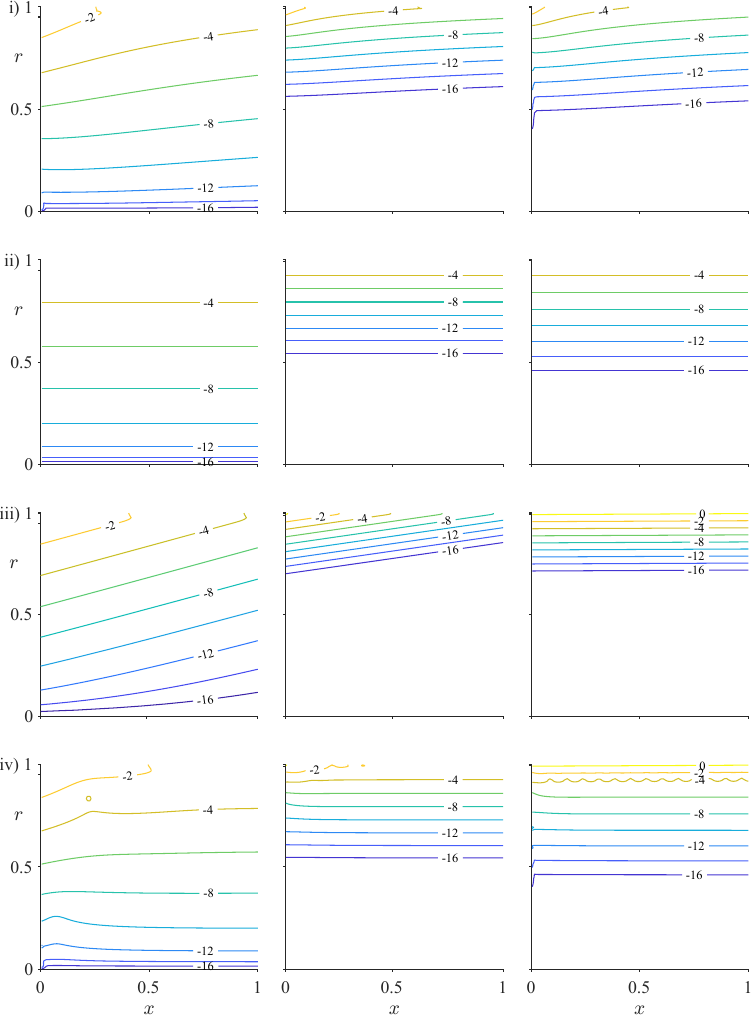}%
\caption{A comparison of the terms that contribute to the critical layer, for $r_0=1-\frac{4h}{5}$. Plotted are the absolute values on a $\log_{10}$ scale.   Left to right: parameter sets~\setA, \setD\ and~\setE. Top to bottom:
(i)~the sum of the three steepest descent contours, $I_{\bk}+I_{r}+I_0$; (ii)~the non-modal $k_0$ pole; (iii)~the contribution of the $k^+$ pole located behind the branch cut; and (iv)~the total contribution from integrating around the critical layer branch cut, obtained by summing (i)--(iii).}%
\label{fig:CLBCCompParams}%
\end{figure}

Figure~\ref{fig:CLBCCompParams}
  illustrates, for three different parameter sets (left to right columns), the differences between the three types of contributions occurring due to the presence of the critical layer branch cut: the sum of the three steepest descent contour integrals (row~(i)); the $k_0$ non-modal pole (row~(ii)); and the $k^+$ modal pole (row~(iii)), which is located below the branch cut for all three parameter sets and therefore does not appear in the modal sum.  The sum of these contributions is plotted on row~(iv) of figure~\ref{fig:CLBCCompParams}.  Comparing the non-modal $k_0$ pole (row~(ii)) to the sum of the three steepest descent integrals (row~(i)), the non-modal $k_0$ pole appears in all three parameter sets to have a small contribution compared to the sum of the three steepest descent integrals for small $x$, although it is comparable and even dominant for larger $x$. For small $x$, the $k^+$ pole's contribution (row~(iii)) is greater than those of the three steepest descent integrals (row~(i)) and the non-modal $k_0$ pole (row~(iii)), particularly near the wall at $r=1$.  However, since the $k^+$ pole decays exponentially in $x$, the non-modal pole will dominate the far-field behaviour of the critical layer branch cut, as can be seen by comparing row~(ii) with row~(iv).
  
\begin{figure}%
    \centering%
\includegraphics{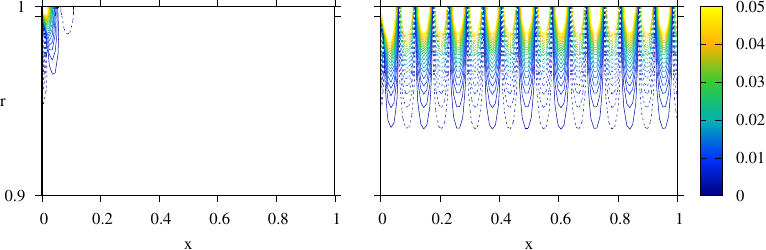}%
    \caption{Plots of the real part of the contribution from integrating around the branch cut ($\Real(p(x,r))$) for parameter set~\setE, excluding any $k^+$ pole located below the branch cut.  Solid lines indicate positive values, dashed lines indicate negative values. (a) $r_0=1-\frac{9h}{10}$; (b) $r_0=1-\frac{3h}{5}$.}%
    \label{fig:CLBCCompr0}%
\end{figure}

  We can further look at how these contributions vary as we adjust the location of the source, shown in figure~\ref{fig:CLBCCompr0}.
The contribution of the non-modal $k_0$ pole is seen to be far smaller for the case when $r_0$ is closest to $1-h$ (figure~\ref{fig:CLBCCompr0}(left)).  Note that there is no non-modal $k_0$ pole when $r_0 \leq 1-h$, in which case the dominant contribution will be from the $k^+$ pole and the steepest descent contours, which in all cases decay to zero.

\begin{figure}%
    \centering%
    \includegraphics{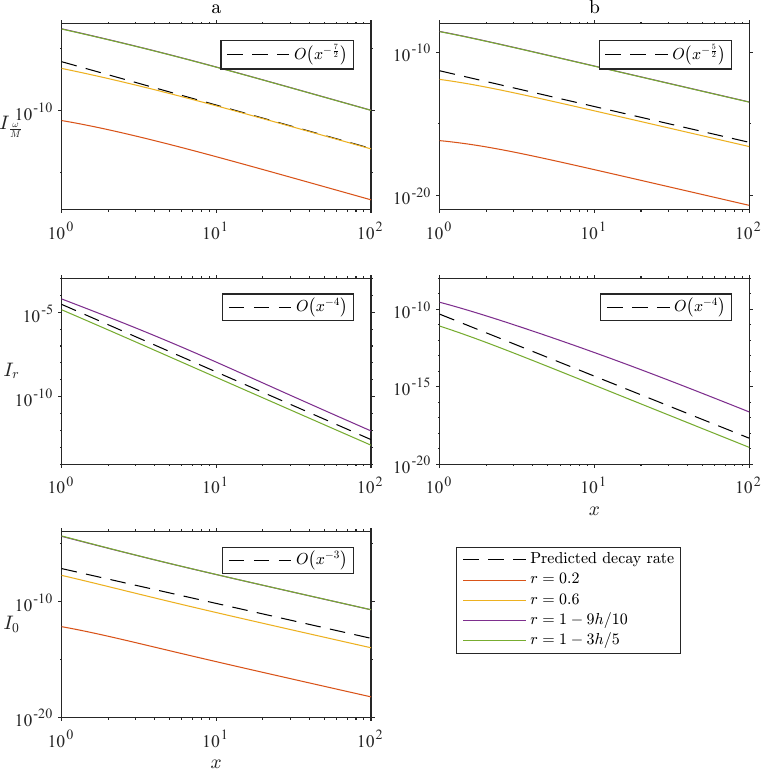}\vspace{-1ex}%
    \caption{Plotted for parameter set~\setA\ are $|I_{\bk}|$ (top), $|I_r|$ (middle), and $|I_0|$ (bottom).  The point source is at $r_0=1-\frac{4h}{5}$ (a) and $r_0=0.4$ (b).  Solid lines correspond to radial locations $r=0.2,0.6,1-\frac{9h}{10}$, and $1-\frac{3h}{5}$.  The dashed line is the predicted far-field rate of decay according to section \ref{sect:Inhomg:Steepest}. Note that for $r=0.2$ and $r=0.6$, $I_r$ is identically zero, since the branch point does not exist. Similarly for $r_0=0.4$ and $I_0$.}
    \label{fig:SteepestRates}%
\end{figure}

Figure~\ref{fig:SteepestRates}
compares the numerically-computed steepest descent integrals with their predicted far-field rates of decay given in section~\ref{sect:Inhomg:Steepest}, and a good agreement is seen in all cases.

\subsection{Full Fourier Inversion}\label{Sect:Nums:FullInv}

\begin{figure}
    \centering%
    \hspace*{\stretch{1.2}}(a)\hspace{\stretch{2}}(b)\hspace*{\stretch{1.5}}\par%
\includegraphics{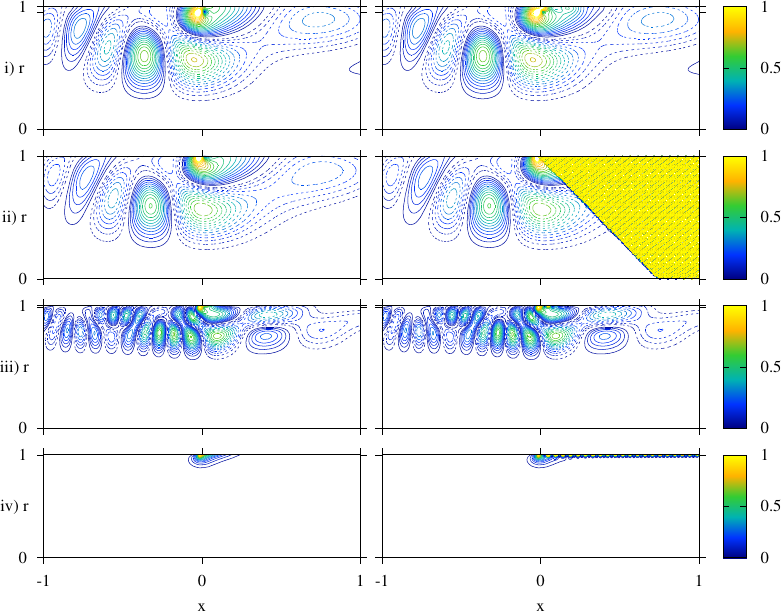}%
    \caption{Plotting the real values of the different contributions.  (a)~just the contribution for the stable modal poles. (b)~the Full Fourier Inversion, which also includes the $k^+$ pole. The parameter sets used from top to bottom are~\setA, \setB, \setD\ and~\setE, with $r_0=1-\frac{4h}{5}$ in each case. In case~\setB, the $k^+$ pole is a convective instability. In cases~\setA, \setD\ and~\setE, the $k^+$ pole is located behind the branch cut.}%
    \label{fig:FullContComp}%
\end{figure}

We now consider the full Fourier inversion, including the contribution from all the modal poles as well as the critical layer branch cut contribution considered above.  Figure \ref{fig:FullContComp}
compares a snapshot in the near-field (for small $x$ values) of the wave field generated by only the stable modal poles (left) with the full solution including the critical layer and any unstable $k^+$ pole (right).  When the $k^+$ pole is a convective instability, it clearly dominates the solution sufficiently far downstream, as it grows exponentially in $x$.  In these near-field plots, the critical layer often appears negligible compared with the modal sum, although in some circumstances it can have a significant effect, as shown by the plots of case~\setE.

\begin{figure}%
    \centering%
    \hspace*{\stretch{1.6}}(a)\hspace{\stretch{2}}(b)\hspace{\stretch{2}}(c)\hspace*{\stretch{1.6}}\par%
\includegraphics{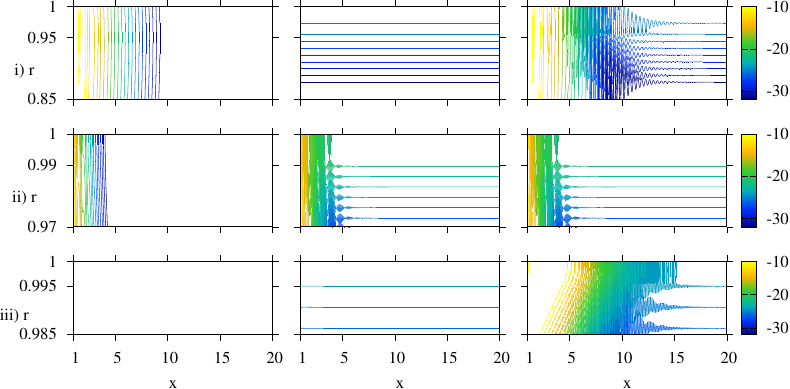}%
    \caption{The absolute value of pressure on a log scale ($10\log_{10}(|p|)$) over a longer range of axial distances downstream of the point source.  (a)~The contribution for the modal poles. (b)~Modal poles plus the three steepest descent contours and the $k_0$ non-modal pole. (c)~The full Fourier inversion, which also includes the $k^+$ pole. The parameter sets used from top to bottom are~\setA, \setD\ and~\setE, with $r_0=1-\frac{4h}{5}$ in each case.  In each case, the $k^+$ pole is located behind the branch cut.  In the bottom left plot, the contribution from the modal poles is too small to be shown (with $10\log_{10}(|p|) < -78$).}%
    \label{fig:FullContCompLongAbs}%
    \vspace{0.5in}%
\end{figure}

In comparison, figure~\ref{fig:FullContCompLongAbs}
shows the behaviour outside the near field for the three stable cases from figure~\ref{fig:FullContComp}, plotting the amplitude of oscillations $|p|$ on a logarithmic scale.  In all cases, since the modal sum decays exponentially, in the far field the dominant contribution is from the critical layer, and this is often true from only one or two radii downstream of the point source.

\begin{figure}
    \centering
\includegraphics{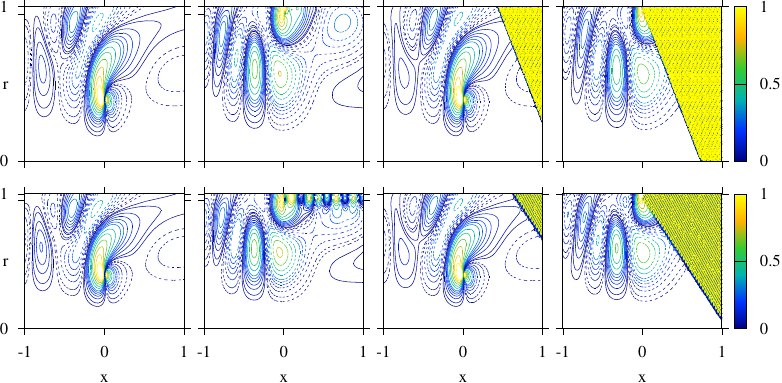}
\caption{Plotting the real values of the full solution for a quadratic boundary layer flow profile (top) and a linear boundary layer flow profile (bottom)~\citep[from][]{brambley2012critical}. From left to right, parameters are: set~\setA\ with $r_0=0.4$; set~\setA\ with $r_0=1-\frac{4h}{5}=0.96$; set~\setB\ with $r_0=0.4$; and set~\setB\ with $r_0=1-\frac{4h}{5}=0.9992$.}
    \label{fig:FullLinComp}
\end{figure}

Figure~\ref{fig:FullLinComp}
compares the wave field generated in a quadratic boundary layer with the wave field in a linear boundary layer profile~\citep[as studied by][]{brambley2012critical}.  The wave fields are reasonably similar, although when the point mass source is within the boundary layer significant differences are seen downstream.  This is related to whether the $k^+$ pole is located above or below the branch cut.  In the quadratic case, the $k^+$ pole always contributes, whether it is behind the branch cut or not, while the $k^-$ is always found above the branch cut and so is not seen to contribute at all.  With the linear boundary layer, instead we find a $k^-$ pole that can be located either above or below the branch cut, while the $k^+$ pole is instead located above in all cases.  The result of this is that the linear boundary layer profile is always found to be convectively unstable, while the quadratic boundary layer profile is only found to be unstable if the boundary layer is sufficiently thin.  Even when both flow profiles give rise to a convective instability, we can see in figure~\ref{fig:FullLinComp} that the growth rate of the instability can be significantly different.

The change in nature of the $k^+$ pole in the quadratic case from convective instability to stable is clearly of significant importance.  We therefore finally consider the variation of the solution as various parameters of interest are varied next.

\clearpage\subsection{Variation of results with changing parameters}\label{Sect:Nums:Stab}

The variation of the acoustic modal sum is relatively well understood, so in this section we concentrate on the variation of the $k^+$ and $k^-$ modal poles as various parameters are varied.  This includes whether $\Imag(k^+)>0$, corresponding to a convective instability, or $\Imag(k^+)<0$, corresponding to a stable modal pole hidden behind the branch cut that none-the-less contributes to the modal sum through the branch cut contribution.  We also consider whether $\Imag(k^-)>0$, meaning the pole is not included, or whether $\Imag(k^-)<0$, in which case the pole is included as part of the contribution of the critical layer branch cut.

\begin{figure}
    \centering%
\includegraphics{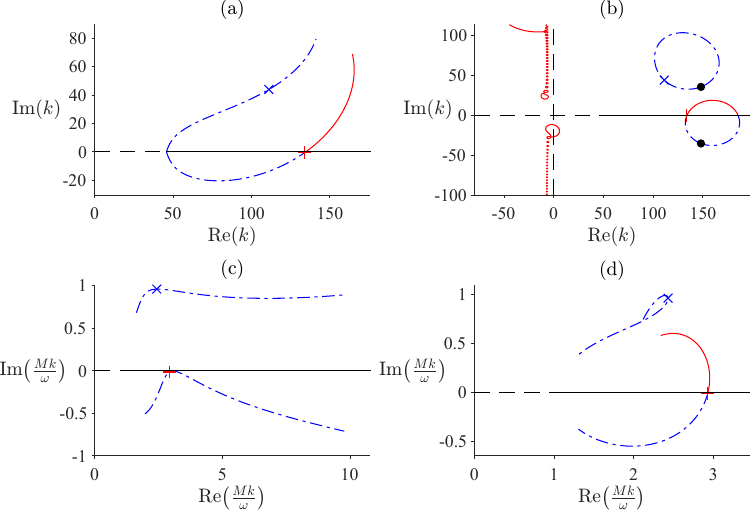}%
\setlength{\unitlength}{0.01\textwidth}\begin{picture}(0,0)
  %
  \qbezier(-60,57)(-65,53)(-70,52)\put(-70,52){\vector(-5,-1){0}}
  \qbezier(-50,47)(-51,44)(-55,42)\put(-55,42){\vector(-2,-1){0}}
  %
  \qbezier(-13,53)(-16,56)(-14,60)\put(-14,60){\vector(1,4){0}}
  \qbezier(-3,53)(2,50)(-2,45)\put(-2,45){\vector(-4,-5){0}}
  %
  \qbezier(-65,12)(-72,15)(-74,12)\put(-74,12){\vector(-2,-3){0}}
  \qbezier(-55,24)(-65,23)(-73,25)\put(-73,25){\vector(-4,1){0}}
  %
  \qbezier(-22,22)(-19,24)(-15,25)\put(-15,25){\vector(4,1){0}}
  \qbezier(-6,10)(0,18)(-5,23)\put(-5,23){\vector(-1,1){0}}
\end{picture}%
    \caption{Motion of the modal poles for parameter set~\setE\ as one parameters is varied (arrows show the motion as the parameter is increased): (a)~varying $h$ in $(0.001,0.5)$; (b)~varying $\Imag(Z)$ in $(-\infty,\infty)$, with a dot showing hard-walled values; (c)~varying $\Real(\omega)$ in $(1,50)$; and (d)~varying $M$ in $(0.06,0.9)$.  Modal positions for parameter set~\setE\ are denoted $+$ ($k^+$) and $\times$ ($k^-$). Dashed lines denote a pole hidden behind the branch cut.  Note that (c) and~(d) use a rescaled $k$ plane in order for the branch cut to remain fixed as $\omega$ or $M$ are varied.}
    \label{fig:kpolehMovement}
\end{figure}
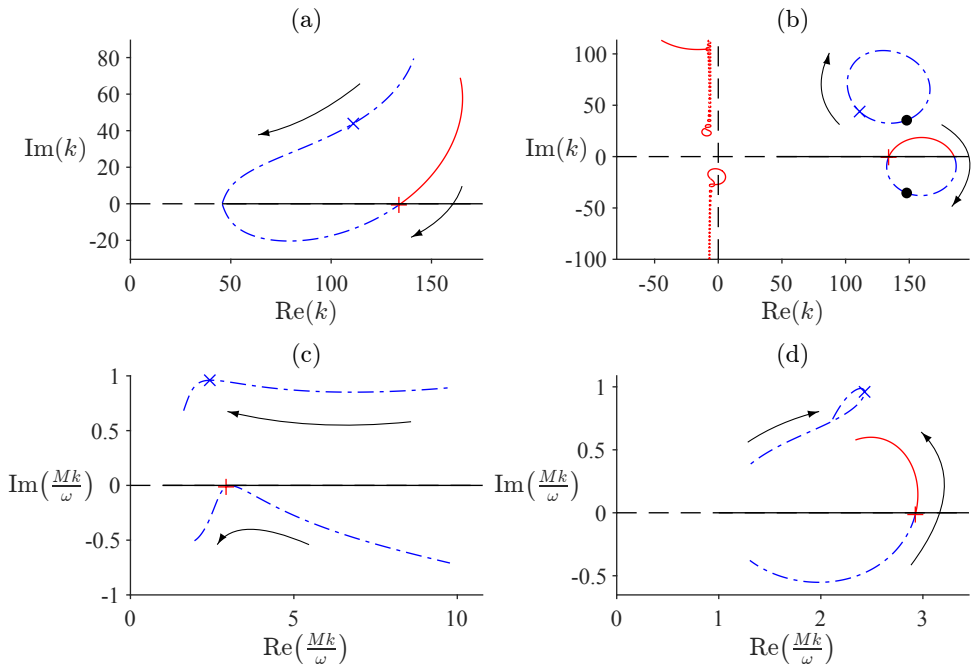

Figure~\ref{fig:kpolehMovement}
illustrates how the $k^+$ and $k^-$ modal poles vary with boundary layer thickness, frequency, impedance and Mach number.  In particular, taking wider boundary layers and lower mean flow velocities  appears to stabilize the $k^+$ pole, moving it to below the branch cut.  In contrast, thinner boundary layers and higher mean flow velocities lead to convective instability.  The value of the impedance is also seen to alter the stability of the solution, with, in this case, a range of values of $\Imag(Z)$ being unstable and nearly hard-walled values of $|\Imag(Z)|\to\infty$ being stable, as is seen from the $k^+$ poles movement in figure~\ref{fig:kpolehMovement}(b).  The variation of stability as the frequency is varied remains unclear, although it appears likely from figure~\ref{fig:kpolehMovement}(c) that, for certain parameters, there would be a finite range of frequencies for which the $k^+$ pole would be unstable, while for frequencies either higher or lower than this range the $k^+$ pole would be stable.  Note also from figure~\ref{fig:kpolehMovement} that, in all cases, the $k^-$ pole is located above the branch cut and so does not contribute either to the modal sum or the critical layer branch cut.

\pagebreak\section{Conclusions}
\label{section:conclusion}

In this work we have considered a cylindrical duct containing a parallel mean flow that is uniform everywhere except within a boundary layer of thickness $h$ near the wall.  Within the boundary layer, which need not be thin, the flow has a quadratic profile and satisfies the non-slip boundary condition at the duct wall, whilst maintaining a $C^1$ continuous flow. For such a flow profile, irrespective of the the thickness of the boundary layer, a solution to the Pridmore-Brown equation has been constructed making use of two Frobenius series expansions, valid for any wave number $k$.  This enables the evaluation of the Greens function of the Pridmore-Brown equation, which is found to consist of a sum of the usual acoustic duct modes plus a non-modal contribution from the critical layer branch cut.  Full source code is provided in the supplementary material to evaluate all solutions presented here.  

In this work, we have aimed to construct the Greens function solutions, which is equivalent to the solution for a point mass source in the linearised Euler equations.  The Greens function is in some sense the general solution, as the solution subject to any forcing can be written as an integral over the Greens function, suitably weighted.  Because of this, any behaviour the equations are capable of must necessarily be demonstrated in the Greens function solution, and so once the behaviour of the Greens function is understood, the equations cannot hold any further surprises.  This is particularly important in this case, considering that the Greens function solution has been shown to include non-modal contributions, such as the critical layer branch cut, which cannot be investigated clearly using other methods such as eigenfunction methods that capture only the modal contributions. 

The Frobenius series method employed here has two particular advantages over other numerical methods to solve the Pridmore-Brown equation~\citep[such as finite differences, e.g.,][]{brambley2011well}.  The first is that the Frobenius series, being a series solution about the critical points of the equation, is at its most accurate near the critical layer singularity found in the Pridmore-Brown equation. This allows for accurate numerical solutions near the critical layer, required for the integration around the critical layer singularity and its associated branch cut to evaluate their effect on the resulting pressure field. Other numerical methods such as finite difference are typically at their least accurate near the critical layer~\citep{brambley2012critical}.  Moreover, the Frobenius series solution makes explicit the branch cut along the critical layer, allowing for analytic continuation of the solution behind the branch cut.  This allows for tracking hydrodynamic instability surface wave modes as they become stable and enter the critical layer (as seen figure~\ref{fig:kpolehMovement}), which makes it significantly easier to track the boundary between stable and unstable behaviour.

An advantage of considering this particular quadratic flow profile is that the origins of the critical layer are on a more solid footing.  For the linear flow profile~\citep{brambley2012critical}, the critical layer was due to the cylindrical geometry of the duct, where as in general the critical layer is due to a non-zero second derivative of the sheared flow profile.  This also allows comparison to previous works, such as that of \citet{swinbanks1975sound} and \citep{felix2007acoustic}. Further, as the quadratic flow profile has a continuous first derivative, we have also been able to investigate the specific case of a point mass source at the boundary between uniform and sheared flow, $r_0=1-h$, and we find this case retains the same behaviour as a point mass source within the region of uniform flow.  In contrast, for the linear flow profile, $r_0\to1-h$ is a singular limit.  We therefore believe the results of the quadratic boundary layer flow profile to be in some ways typical of solutions for a general boundary layer profile.

The final solution for the Green's function for a point mass source was found to consist of a number of contributions. This solution is dominated, both upstream and, in the near field, downstream too, by the sum of modal poles.  The modal poles, including acoustic and surface modes, are well known, and are typically used in mode-matching numerical methods.  One complication found here to the surface modes is that a particular surface mode, here labelled $k^+$, is found to sometimes disappear behind the critical layer branch cut (or, in other words, into the continuous spectrum).  The contribution of this mode is not lost, however, and is in effect added to the critical layer contribution.  In general, the modal poles present difficulty only in establishing which poles contribute upstream $(x<0)$ or downstream $(x>0)$ of the source. This can be established through application of the Briggs--Bers criterion, as summarised in section~\ref{sect:Inhomg:Inversion}.

The effect of the critical layer, the focus of this work, always contributes downstream of the source, and is the dominant contribution to the far-field pressure downstream of the point mass source.  This contribution, which results from integrating the Fourier inversion contour around the critical layer branch cut, may be viewed in three parts. The first contribution is from the $k_0$ non-modal pole, only present when $r_0>1-h$, which does not decay with distance from the point source and is therefore dominant in the far-field downstream of the source.  This contribution is similar to that described in the linear flow profile case~\citep{brambley2012critical}, and may be interpreted similarly as a hydrodynamic vorticity wave generated from the point mass source interacting with the sheared mean flow, travelling downstream with phase velocity equal to the local fluid velocity $U(r_0)$. The second contribution to the critical layer is from the steepest-descent non-oscillatory integrals $I_{\bk}$, $I_r$ and $I_0$, which are the results of accounting for the branch points coming from the critical points of the Pridmore-Brown equation.  These contributions have a phase speed equal to, respectively, the uniform flow speed $M$, $U(r)$, and $U(r_0)$, and decay algebraically in the far-field downstream of the point source.  The final contribution is from any modal pole that is hiding behind the branch cut, such as from a $k^+$ surface wave mode that has stabilized by moving into the critical layer branch cut from above.  These poles, while looking very much like ordinary modal poles, are not able to be found by traditional numerical methods, as they require analytically continuing behind the critical layer branch cut.  While these poles decay exponentially with distance downstream of the point source, their decay rate may be slower than any other acoustic modal pole, depending on the parameters used, and so may still be significant in the far-field; this was seen for parameter set~\setE\ in figure~\ref{fig:FullContCompLongAbs}.

The $k^+$ modal pole may be present as a hydrodynamic instability surface wave, or as a stable mode included within the critical layer branch cut contribution.  Interestingly, in the linear flow profile case~\citep{brambley2012critical}, this mode was always an instability and was never hidden behind the critical layer branch cut.  From the results of figure~\ref{fig:kpolehMovement}, we expect that this mode is stable for quadratic flow boundary layer profiles when the boundary layer is sufficiently thick or the flow speed is sufficiently slow, although the specific stability boundary also depends on the impedance $Z$ and frequency $\omega$.

For the linear flow profile boundary layer~\citep{brambley2012critical}, a $k^-$ pole was found below and behind the critical layer branch cut that contributed to the critical layer.  For the quadratic flow profile boundary layer here, this $k^-$ pole is always found to be above the critical layer branch cut, and so never contributes.  We believe that this $k^-$ pole was an artifact of the unphysical linear boundary layer profile, although we have no direct way of demonstrating this.  Incidentally, for the linear flow profile boundary layer, \citet{brambley2012critical} argued that the $k^+$ pole could never be behind the critical layer branch cut, as this would cause an unphysical discontinuity in the final solution; in fact, it is found here that when the $k^+$ pole is behind the branch cut, the unphysical discontinuity in the $k^+$ pole contribution is exactly cancelled by the $I_r$ steepest descent contour contribution, resulting in a continuous solution.

The various decay rates of the components of the critical layer have previously been predicted by~\citet{swinbanks1975sound} and~\citet{brambley2012critical}; and a summary can be found in table~\ref{tab:results}. \Citet{swinbanks1975sound} only considered the contribution from waves with phase velocity $U(r_0)$, which are only present for a point mass source within the region of sheared flow, $r_0 > 1-h$.  \Citet{swinbanks1975sound} predicted these to behave as a constant amplitude plus a decay as $O(x^{-3})$ in the far field.  \Citet{brambley2012critical} found the same result, despite \citeauthor{swinbanks1975sound} considering a two dimensional flow in a rectilinear duct with an arbitrary flow profile and \citeauthor{brambley2012critical} considering only a constant-then-linear flow profile in a three dimensional cylindrical duct;  in particular \citeauthor{swinbanks1975sound} emphasised the importance of the non-zero second derivative of the mean flow profile, which is identically zero for a constant-then-linear flow profile. As a result, it would not have been unsurprising if these results were different.  Here, the same result is again found, with the constant amplitude coming from the $k_0$ non-modal pole and the algebraic decay coming from the $I_0$ integral.  This shows that this agreement is not by coincidence. For the critical layer contribution that propagates with phase velocity $U(r)$ when $r$ is within the boundary layer, we also find here the same result given by \citet{brambley2012critical} of an $O(x^{-4})$ far-field decay.

The critical layer also contributes a term with phase velocity equal to the uniform flow velocity $M$, which is always present, and which dominates the other critical layer contributions in the far field whenever the point mass source is in the uniform flow region, $r_0<1-h$. The amplitude of this term can decay at two different rates depending on the location of the source. When the source is within the uniform flow a decay rate of $O(x^{-\frac{5}{2}})$ is found, while when the source is within the sheared flow we instead have a faster rate of decay of $O(x^{-\frac{7}{2}})$. These results differ from those found by \citet{brambley2012critical} in the linear flow profile boundary layer case, despite corresponding to the same physical behaviour. This difference may be understood as result of both the difference in the overall shape of the flow profile, and the importance of the second derivative of the flow.  Indeed, we conjecture that these scalings will differ depending on the flow profile within the boundary layer, and an example discussion of this for $n$-polynomial flow profiles is given in appendix~\ref{appendix:n-polynomial}.

In most aeroacoustic analyses, particularly those involving mode matching, the critical layer is either implicitly or explicitly ignored.  The work here suggests that this may be valid in the near-field provided not all acoustic modes are cut-off, although even in the near-field the critical layer can be dominant if all acoustic modes are cut-off, as shown in figure~\ref{fig:FullContComp} for parameter set~\setE.  However, it is certainly not valid to ignore the critical layer downstream in the far-field, when the critical layer will be the dominant contribution.  Moreover, without considering the critical layer, it would not be apparent whether a barely-stable hydrodynamic surface wave is present only just hidden behind the critical layer branch cut (or, in other words, within the continuous spectrum).

There are a number of possible avenues for further investigation following on from this work.  One of practical importance concerns whether the hydrodynamic surface wave $k^+$ can be accurately predicted using a surface wave dispersion relation~\citep[e.g.][]{brambley2013surface}, especially when the $k^+$ pole is located behind the critical layer branch cut; our experience in this work has been that it cannot, at least with the simplified surface wave dispersion relations that assume a thin boundary layer with a linear flow profile, although more complicated surface wave dispersion relations may prove more accurate.  Another possibility for further investigation is to consider parameters on the stability boundary when the $k^+$ pole is neutrally stable.  In this case, the $k^+$ pole is exactly located on the critical layer branch cut, and there would also exist a value of $r_0$ for which the non-modal $k_0$ pole and the $k^+$ pole overlap; this case has been explicitly excluded here.  While this may seem a rather contrived case, a distributed sound source would correspond to an integral of source strength over all values of $r_0$, and so $k_0$ and $k^+$ coinciding could be expected to occur for any parameters leading to exact neutral stability.  One could also extend this problem to a non-constant mean density and sound speed making use of equation \eqref{PridmoreBrownFull}. For a given mean density profile $\rho_0(r)$ and sound speed profile $c_0(r)$, one could still construct a solution to the resulting Pridmore-Brown equation, taking careful notice of the potentially complex roots of $c_0(r)$. It would be possible to construct a solution using the Frobenius series solutions still so long as these are not double roots or of higher order, and $\frac{\rho_0^\prime(r)}{\rho(r)}$ has at most regular singularities in the complex r plane. Except in the case where these singularities occur at $r=r_c^+$ then the critical layer branch cut will still occur in identical form to that seen in our work, although the resulting scaling in the various limits seen above may vary. When retrieving these, the work given here would provide a suitable outline for the approach to be taken.  Finally, the critical layer may be regularized by considering either viscosity or weak nonlinearity, and it would be interesting to investigate how the results presented here are recovered in the inviscid or small-amplitude limits.  In particular, for viscous thin boundary layers, the critical layer is recovered as a caustic in the high-frequency limit~\citep{brambley2011acoustic}.

\begin{acknowledgements}
\textbf{Supplementary material.} The Matlab source code used to produce the data plotted here is available at \url{https://doi.org/10.1017/jfm.2022.753}.

\textbf{Acknowledgements.} MJK was supported in this work through the University of Warwick MASDOC Doctoral Training Centre, and gratefully acknowledges their support.
EJB gratefully acknowledges the support of a Royal Society University Research Fellowship (UF150695 and RGF\textbackslash EA\textbackslash 180284).
RL was supported in this work through a research internship funded by the Royal Society (RGF\textbackslash EA\textbackslash 180284), and would also like to thank the CAPES~Foundation, Ministry of Education of Brazil for the award of a BRAFITEC scholarship.
The contribution of MR was supported by an EPSRC UROP undergraduate research summer internship (2015, DAMTP, University of Cambridge).

\textbf{Declaration of Interests.} The authors report no conflict of interest.
\end{acknowledgements}

\appendix

\pagebreak
\section{Frobenius series solutions to the Pridmore-Brown equation with a quadratic mean flow profile}
\label{appendix:frobenius}

In this appendix, we use a Frobenius expansion method to solve the homogeneous Pridmore-Brown equation~\eqref{HomogPridmoreBrown},
\begin{equation}\label{appendix:HomogPridmoreBrown}
    \tp^{\prime\prime}+\lr{\frac{2k U^\prime}{\omega-U(r)k}+\frac{1}{r}}\tp^\prime+\lr{(\omega-U(r)k)^2-k^2-\frac{m^2}{r^2}}\tp=0,
\end{equation}
for the flow profile~\eqref{FlowProfile},
\begin{equation}\label{appendix:FlowProfile}
    U(r)=\begin{cases}M & 0\leq r\leq 1-h  \\
    M(1-(1-\frac{1-r}{h})^2) & 1-h\leq r \leq 1,\end{cases}
\end{equation}
in the quadratic flow region $1-h\leq r \leq 1$.  The Pridmore-Brown equation~\eqref{appendix:HomogPridmoreBrown} has regular singularities at $r=0$ and at $r=r_c$, where $\omega - U(r_c)k=0$.  For the quadratic flow profile~\eqref{appendix:FlowProfile}, the solutions of $\omega-U(r_c)k=0$ are given by~\eqref{rcplusminus},
\begin{align}\label{appendix:rcplusminus}
    r_c^\pm&=1-h\pm Q, &  \text{where}\quad Q&=h\sqrt{1-\frac{\omega}{Mk}}.
\end{align}
This results in the Pridmore-Brown equation in the quadratic flow region $1-h\leq r\leq 1$ being given by
\begin{equation}\label{appendix:ShearedHomogPridmoreBrown}
    \tp^{\prime\prime}+\lr{\frac{2}{r-r_c^+}+\frac{2}{r-r_c^-}+\frac{1}{r}}\tp^\prime+\lr{\frac{M^2k^2}{h^4}(r-r_c^+)^2(r-r_c^-)^2-k^2-\frac{m^2}{r^2}}\tp=0.
\end{equation}
We choose $\Real(Q)\geq 0$ and consider the Frobenius expansion about $r = r_c^+$.
\subsection{Frobenius expansion about \texorpdfstring{$r=r_c^+$}{r=rc+}}
\label{appendix:frobeniusc}

Following~\citet{brambley2012critical}, we propose a Frobenius expansion about the regular singularity $r_c^+$,
\begin{align}\label{appendix:FrobeniusMethod}
    \tp(r)&=\sum_{n=0}^\infty a_n(r-r_c^+)^{n+\sigma} &
    &\text{with}\qquad a_0 \neq 0.
\end{align}
We substitute~\eqref{appendix:FrobeniusMethod} into~\eqref{appendix:ShearedHomogPridmoreBrown} and expand using a Laurent series.  Specifying that $a_0 \neq 0$ results in the requirement that $\sigma(\sigma-3)=0$.  By Fuchs theorem~\citep{teschl2012ordinary}, this gives a pair of linearly independent solutions of the form
\begin{subequations}\label{appendix:ShearSolnCrit}\begin{align}
    \tp_{c1}(r)&=\sum_{n=0}^\infty a_n (r-r_c^+)^{n+3}, \\
    \tp_{c2}(r)&=A\tp_{c1}(r)\log(r-r_c^+)+\sum_{n=0}^\infty b_n(r-r_c^+)^n.
\label{appendix:ShearSolnCrit2}
\end{align}
The coefficients $a_n$ and $b_n$ are then given by setting the remaining terms of the Laurent expansion of~\eqref{appendix:ShearedHomogPridmoreBrown} to zero, resulting in the recurrence relation
\begin{align}
   a_n&=\frac{1}{n(n+3)}\Bigg[ k^2a_{n-2}-\frac{k^2M^2}{h^4}\lr{a_{n-6}+ 4Qa_{n-5}+4Q^2a_{n-4}}
    \notag\\&\qquad\qquad\qquad
    -\sum_{j=0}^{n-1}(-1)^j(n+2-j)a_{n-1-j}\lr{\frac{1}{(r_c^+)^{j+1}}-\frac{2}{(2Q)^{j+1}}}
    \notag\\&\qquad\qquad\qquad
     +m^2\sum_{j=0}^{n-2}\frac{(-1)^j}{(r_c^+)^{j+2}}(j+1)a_{n-2-j} \Bigg],
\label{appendix:an}
\\
    b_n&=-\frac{1}{n(n-3)}\Bigg[ A\Bigg(\!(2n-3)a_{n-3}+\sum_{j=0}^{n-4}a_{n-4-j}(-1)^j\!\lr{\frac{1}{(r_c^+)^{j+1}}-\frac{2}{(2Q)^{j+1}}}\!\!\Bigg)
    \notag\\&\qquad\qquad\qquad
    -k^2b_{n-2}+\frac{k^2M^2}{h^4}\lr{b_{n-6}+4Qb_{n-5}+4Q^2b_{n-4}}
     \notag\\&\qquad\qquad\qquad
    +\sum_{j=0}^{n-1}(-1)^j(n-1-j)b_{n-1-j}\!\lr{\frac{1}{(r_c^+)^{j+1}}-\frac{2}{(2Q)^{j+1}}}
    \notag\\&\qquad\qquad\qquad
    -m^2\sum_{j=0}^{n-2}\frac{(-1)^j}{(r_c^+)^{j+2}}(j+1)b_{n-2-j} \Bigg],
\label{appendix:bn}
\end{align}
\end{subequations}
where we take $a_n=b_n=0$ for $n<0$.  Requiring $a_0 = b_0 = 1$, we then find that
\begin{align}
    b_1 &= 0, &
    b_2 = &-\frac{1}{2}\!\lr{\!k^2+\lr{\frac{m}{r_c^+}}^{\!\!2}}\!,
\end{align}
and that $b_3$ is arbitrary, so we choose $b_3=0$.  However, for the recurrence relation involving $b_3$ on the left hand side to hold, we also require that $A$ is chosen to be
\begin{equation}\label{appendix:equ:A}
    A=-\frac{1}{3}\lr{\frac{1}{Q}-\frac{1}{r_c^+}}\!\lr{\!k^2 + \lr{\frac{m}{r_c^{+}}}^{\!2}} - \frac{2m^2}{3r_c^{+3}}.
    \end{equation}
Here, the notation $\tp_{c1}$ and $\tp_{c2}$ denotes that these are two linearly independent solutions for $\tp$ about the critical point $r_c^+$.

\begin{figure}
    \centering
\begin{tikzpicture}[scale=1.2]
\draw (0,5) rectangle (5,0);
\draw (1,1.25) --(1.25,1.25) node[anchor= north east] {$0$};
\draw[thick,->] (1,1.25) -- (4,1.25) node[anchor= west] {$\Real(r)$};
\draw[thick,->] (1.25,1) -- (1.25,4) node[anchor= south] {$\Imag(r)$};
\draw[thick] (2.5,1.25+0.1) -- (2.5,1.25-0.1) node[anchor= north] {$1-h$};
\draw[thick] (3.75,1.25+0.1) -- (3.75,1.25-0.1) node[anchor= north] {$1$};
\draw [pattern=north west lines, pattern color=blue] (2.5+0.5,1.25+0.25) circle (1.11803398875cm);
\fill (2.5+0.5,1.25+0.25) circle (0.05cm) node[anchor= south west] {$r_c^+$};
\fill (2.5-0.5,1.25-0.25) circle (0.05cm) node[anchor= north east] {$r_c^-$};
\draw[ultra thick] (2.5,1.25) -- (3.75,1.25);
\node[anchor = north east] at (4.75,4.75) {(a)};

\draw (0+6,5) rectangle (5+6,0);
\draw (1+6,1.25) --(1.25+6,1.25) node[anchor= north east] {$0$};
\draw[thick,->] (1+6,1.25) -- (4+6,1.25) node[anchor= west] {$\Real(r)$};
\draw[thick,->] (1.25+6,1) -- (1.25+6,4) node[anchor= south] {$\Imag(r)$};
\draw[thick] (2.5+6,1.25-0.1) -- (2.5+6,1.25+0.1);
\draw[thick] (3.75+6,1.25+0.1) -- (3.75+6,1.25-0.1) node[anchor= north] {$1$};
\draw [pattern=north west lines, pattern color=blue] (2.5+0.25+6,1.25+0.125) circle (0.5*1.11803398875cm);
\fill (2.5+6+0.25,1.25+0.125) circle (0.05cm) node[anchor= south west] {$r_c^+$};
\fill (2.5+6-0.25,1.25-0.125) circle (0.05cm) node[anchor= north east] {$r_c^-$};
\draw[ultra thick] (2.5+6,1.25) -- (3.75+6,1.25);
\node[anchor = north east] at (10.75,4.75) {(b)};
\end{tikzpicture}
    \caption{Schematic of possible locations of the $r_c^\pm$ critical points in the complex $r$-plane.  (a)~The radius of convergence of the expansion about $r_c^+$ covers the region of interest $r\in[1-h,1]$.  (b)~The radius of convergence about $r_c^+$ is insufficient to cover $r\in[1-h,1]$.}
    \label{appendix:figRad1}
\end{figure}
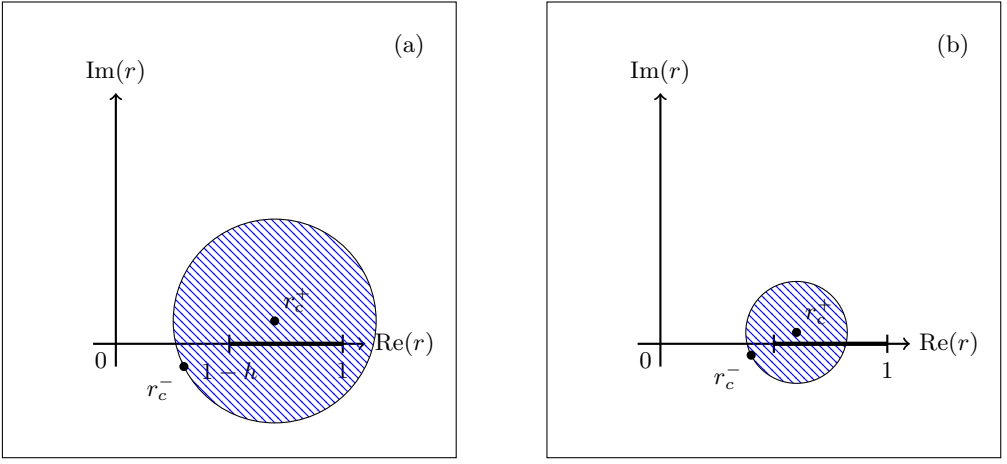

The Frobenius series solutions~\eqref{appendix:ShearSolnCrit} are limited by a radius of convergence, in that the series converge if $|r-r_c^+| < R$ for some radius of convergence $R$. Following from Fuchs Theorem \citep[Theorem 4.5]{teschl2012ordinary} this $R$ is the distance between $r_c^+$ and the next nearest singularity of the Pridmore-Brown equation, which is either at $r=0$ or at $r = r_c^-$, and hence
\begin{equation}
R = \min\big\{|1-h+Q|, 2|Q|\big\}.
\end{equation}
The choice of $r_c^+$ as the singularity to expand around means that this expansion maximizes the region of $[1-h,1]$ contained within the radius of convergence.  This is shown schematically in figure~\ref{appendix:figRad1}.
It can be observed that these solutions are not always valid for all of $r\in[1-h,1]$.  In particular, in the case of $k\to\bk$ we observe $r_c^\pm\to(1-h)$ and the radius of convergence $R \to 0$.

\subsection{Frobenius expansion about \texorpdfstring{$r=1$}{r=1}}
\label{appendix:frobenius1}

In order to cover the remainder of the domain $[1-h,1]$, we construct a second series solution about $r=1$,
\begin{align}
    \tp_{11}(r)&=\sum_{n=0}^\infty \alpha_n (r-1)^{(n+1)}, & \tp_{12}(r)&=\sum_{n=0}^\infty \beta_n (r-1)^n.
\end{align}
Specifying that $\alpha_0 = \beta_0 = 1$, this results in the recurrence relation
\begin{subequations}\begin{align}
    \alpha_n=-\frac{1}{n(n+1)}\Bigg[&-k^2\alpha_{n-2}+\frac{k^2 M^2}{h^4}\bigg(\alpha_{n-6}+4h\alpha_{n-5}+2(3h^2-Q^2)\alpha_{n-4}
    \notag\\[-1em] & \qquad \qquad \qquad \qquad \qquad +4h(h^2-Q^2)\alpha_{n-3}+(h^2-Q^2)^2\alpha_{n-2}\bigg)
    \notag\\[-1ex] &
    +\sum_{j=0}^{n-1}(-1)^j(n-j)\alpha_{n-j-1}\lr{1-\frac{2}{(h+Q)^{j+1}}-\frac{2}{(h-Q)^{j+1}}}
    \notag\\ &
    -m^2\sum_{j=0}^{n-2}(-1)^j(j+1)\alpha_{n-j-2}\Bigg]. 
\label{appendix:alphan}\\
    \beta_n=-\frac{1}{n(n-1)}\Bigg[&-k^2\beta_{n-2}+\frac{k^2 M^2}{h^4}\bigg(\beta_{n-6}+4h\beta_{n-5}+2(3h^2-Q^2)\beta_{n-4}
    \notag\\[-1em] & \qquad \qquad \qquad \qquad \qquad +4h(h^2-Q^2)\beta_{n-3}+(h^2-Q^2)^2\beta_{n-2}\bigg)
    \notag\\[-1ex] &
    +\sum_{j=0}^{n-1}(-1)^j(n-j-1)\beta_{n-j-1}\lr{1-\frac{2}{(h+Q)^{j+1}}-\frac{2}{(h-Q)^{j+1}}}
    \notag\\ &
    -m^2\sum_{j=0}^{n-2}(-1)^j(j+1)\beta_{n-j-2}\Bigg].
\label{appendix:betan}
\end{align}\end{subequations}
with $\alpha_n=\beta_n=0$ for $n<0$.
These solutions are labelled $\tp_{11}$ and $\tp_{12}$ to indicate they are two linearly independent solutions to $\tp$ expanded about the point $r=1$.

\subsection{A homogeneous solution valid across \texorpdfstring{$[1-h,1]$}{[1-h,1]}}
\label{appendix:frobenius-full}

\begin{figure}
    \centering
\begin{tikzpicture}[scale=1.2]
\draw (0+6,5) rectangle (5+6,0);
\draw (1+6,1.25+1) --(1.25+6,1.25+1) node[anchor= north east] {$0$};
\draw[thick,->] (1+6,1.25+1) -- (4+6,1.25+1) node[anchor= west] {$\Real(r)$};
\draw[thick,->] (1.25+6,1) -- (1.25+6,4) node[anchor= south] {$\Imag(r)$};
\draw[thick] (2.5+6,1.25-0.1+1) -- (2.5+6,1.25+0.1+1);
\draw[thick] (3.75+6,1.25+0.1+1) -- (3.75+6,1.25-0.1+1) node[anchor= north] {$1$};
\draw [pattern=north west lines, pattern color=blue] (2.5+0.25+6,1.25+0.125+1) circle (0.5*1.11803398875cm);
\draw [pattern=north east lines, pattern color=red] (3.75+6,1.25+1) circle (1.00778221854cm);

\fill (2.5+6+0.25,1.25+0.125+1) circle (0.05cm) node[anchor= south west] {$r_c^+$};
\fill (2.5+6-0.25,1.25-0.125+1) circle (0.05cm) node[anchor= north east] {$r_c^-$};

\draw[ultra thick] (2.5+6,1.25+1) -- (3.75+6,1.25+1);

\end{tikzpicture}
    \caption{As for figure~\ref{appendix:figRad1}(b) with the radius of convergence for $\tp_{11}$ and $\tp_{12}$ added.}
    \label{appendix:figRad2}
\end{figure}
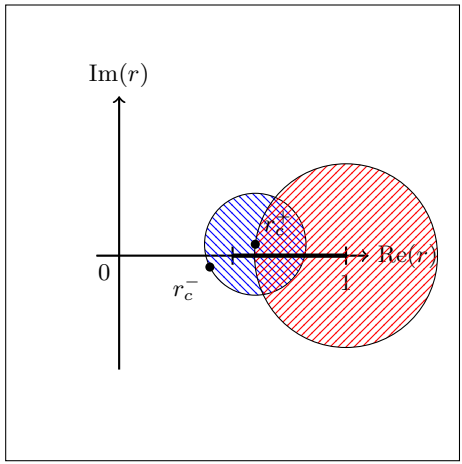

We now construct solutions to the homogeneous Pridmore-Brown equation $\tp_1(r)$ and $\tp_2(r)$ such that they are valid across the whole of $[1-h,1]$.  We set
\begin{subequations}\label{appendix:ShearSolnFull}\begin{align}\label{appendix:ShearSolnFull1}
    \tp_1(r)&=\begin{cases}
    \tp_{c1}(r) & |r - r_c^+| < R \\
    A_1\tp_{11}(r)+B_1\tp_{12}(r) & \text{otherwise}
    \end{cases}\\ 
    \label{appendix:ShearSolnFull2} 
    \tp_2(r)&=\begin{cases}
    \tp_{c2}(r) & |r - r_c^+| < R \\
    A_2\tp_{11}(r)+B_2\tp_{12}(r) & \text{otherwise}
    \end{cases}
\end{align}\end{subequations}
First of all, note that these expansions are sufficient for a uniformly-valid expansion, as sketched in figure~\ref{appendix:figRad2}.
Note also from figure~\ref{appendix:figRad2} that the regions of convergence of the $\tp_{c}$ solutions and the $\tp_{1}$ solutions always overlap (except when $k=\bk$, which we exclude here).  For any real $\rmatch > \Real(r_c^+)$ contained within both regions of convergence, we may find the coefficients $A_1$, $A_2$, $B_1$ and $B_2$ are found by requiring continuity and continuous derivatives at $r = \rmatch$:
\begin{equation}
    \left[\begin{array}{cc}
        A_1 & A_2  \\
        B_1 & B_2
    \end{array}\right]=\left[\begin{array}{cc}
        \tp_{11}(\rmatch) & \tp_{12}(\rmatch)  \\
        \tp_{11}^\prime(\rmatch) & \tp_{12}^\prime(\rmatch)
    \end{array}\right]^{-1}\left[\begin{array}{cc}
        \tp_{c1}(\rmatch) & \tp_{c2}(\rmatch)  \\
        \tp_{c1}^\prime(\rmatch) & \tp_{c2}^\prime(\rmatch)
    \end{array}\right]
\end{equation}
These coefficients $A_1,B_1,A_2$ and $B_2$ are independent of the specific choice of $\rmatch$, and the resulting solutions $\tp_1$ and $\tp_2$ have not only $C^1$ continuity but $C^\infty$, since both are solutions to the Pridmore-Brown equation.  In effect, $\tp_1$ analytically continues $\tp_{c1}$ beyond its radius of convergence, and similarly $\tp_2$ analytically continues $\tp_{c2}$.

As described in~\eqref{JumpP2}, there is a jump in $\tp_{c2}$ across the critical layer branch cut due to the $\log$ term in~\eqref{appendix:ShearSolnCrit2}.  If the radius of convergence $R$ is sufficiently large that $r=1$ is within the radius of convergence, then no matching coefficients are needed, and this jump in $\tp_{c2}$ obviously carries through to $\tp_2$.  In the other case, that $R$ is sufficiently small that matching is needed, it follows that $\rmatch < 1$.  In this case, there is no jump in the matching coefficients $A_1$, $A_2$, $B_1$ and $B_2$ as $\rmatch > \Real(r_c^+)$, and hence
\begin{equation}\label{appendix:jumpP2}
    \Delta\tp_2(r)=-2\upi\mi A \tp_1(r) H(r_c^+-r).
\end{equation}
This is analogous to the jump in $\tp_2$ given in~\eqref{JumpP2}, and shows that the jump in $\tp_{c2}$ carries through the analytic continuation, as might have been expected a priori.

\subsection{The Wronskian of \texorpdfstring{$\tp_1$}{p1} and \texorpdfstring{$\tp_2$}{p2}}
\label{appendix:Sect:Wr}

We define the Wronskian of $\tp_1$ and $\tp_2$ to be
\begin{equation}\label{appendix:Wrontpform1}
    W(r) = \mathcal{W}(\tp_1,\tp_2;r)=\tp_1(r)\tp_2^\prime(r)-\tp_{2}(r)\tp_1^\prime(r).
\end{equation}
Since $\tp_1$ and $\tp_2$ are solutions to the homogeneous Pridmore-Brown equation~\eqref{appendix:HomogPridmoreBrown}, we have that
\begin{align}\label{appendix:Wrontpform2}
   W'+\!\lr{\frac{2k U^\prime}{\omega-Uk}+\frac{1}{r}}\!W&=0 &
&\Rightarrow&
    W(r)&\propto \frac{(r-r_c^+)^2(r-r_c^-)^2}{r}.
\end{align}
By considering the behaviour of $\tp_1$ and $\tp_2$ as $r\to r_c^+$, we find that $W(r) = -3(r-r_c^+)^2 + O\big((r-r_c^+)^4\big)$, and so the constant of proportionality can be found, giving
\begin{equation}\label{appendix:Wr}
    W(r)= -\frac{3}{4}\frac{r_c^+(r-r_c^+)^2(r-r_c^-)^2}{rQ^2}.
\end{equation}

\section{The jump in \texorpdfstring{$\tG$}{G} across the critical layer branch cut}
\label{appendix:DeltaGq}
In this appendix, we split the jump in $\tG$ across the critical layer branch cut, $\Delta\tG$, into its various components about the three possible branch points $\bk$, $k_0$ and $k_r$.  For this reason, we restrict attention to $k \in[\bk,\infty)$, that is, to $k$ on the critical layer branch cut.  In this case, $r_c^+(k)\in[1\!-\!h,1)$, and $r_c^+(k)$ is an increasing function of $k$.  Recall from~\eqref{JumptG} that
\begin{align}\label{appendix:JumptG}
\Delta\tG=&-\frac{\omega-U(r^*)k}{2\I\pi r^*W(r^*)}
\frac{1}{C^-_1\dg_2-\cg_2D_1+2\mi\upi AD_1\dg_2}\\
&\quad\times \left[\frac{2\mi\upi A D_1\dg_2\psi^-_1(\rl)\psi^-_2(\rg)}{C_1^-\dg_2-\cg_2D_1}-\psi^-_1(\rl)\Delta\psi_2(\rg)-\Delta\psi_1(\rl)\psi^-_2(\rg)-\Delta\psi_1(\rl)\Delta\psi_2(\rg)\right]\notag,
\end{align}
with $\Delta\psi_1$ and $\Delta\psi_2$ given in~\eqref{jumptpsi1andpsi2} as
\begin{subequations}\label{Appendix:jumptpsi1andpsi2}\begin{align}
    \Delta\psi_1(r)&=\begin{cases}
    0 &  r < r_c^+\\
    2\mi\upi AD_1\tp_1 &  r \geq r_c^+,
    \end{cases}\\
    \Delta\psi_2(r)&=\begin{cases}
    \Delta\cl_2H_m^{(1)}(\alpha r)+\Delta\dl_2H_m^{(2)}(\alpha r) & 0\leq r\leq 1-h \\
    -2\mi\upi A \tp_1(r)\dg_2 & 1-h\leq r \leq r_c^+\\
    0 & r_c^+< r \leq 1.
    \end{cases}
\end{align}\end{subequations}
Note that since $\rl<\rg$ it must be that $\Delta\psi_1(\rl)\Delta\psi_2(\rg)=0$ in all cases.

When $r,r_0<1-h$ then for any $k$ on the branch cut we have that $\Delta\psi_1 =0$ and $\Delta\psi_2 \neq 0$.  This means that we have the same formula for $\Delta\tG$ for any $k$ on the branch cut in this case, so that $\bk$ is the only branch point of $\Delta\tG$.  Hence, we write $\Delta\tG = \Delta\tG_{\bk}$, where
\begin{align}
\Delta\tG_{\bk}=&-\frac{\omega-U(r^*)k}{2\I\pi r^*W(r^*)}
\frac{1}{C^-_1\dg_2-\cg_2D_1+2\mi\upi AD_1\dg_2} \notag \\
&\quad\times \left[\frac{2\mi\upi A D_1\dg_2\psi^-_1(\rl)\psi^-_2(\rg)}{C_1^-\dg_2-\cg_2D_1}-\psi^-_1(\rl)\Delta\psi_2(\rg)\right]\notag, \\
=&-\frac{\omega-U(r^*)k}{r^*W(r^*)}
\frac{1}{C^-_1\dg_2-\cg_2D_1+2\mi\upi AD_1\dg_2} \frac{ A\dg_2^2\psi^-_1(\rl)\psi^-_1(\rg)}{C_1^-\dg_2-\cg_2D_1}
\end{align}

When $\rl<1-h<\rg$ then the formula for $\Delta\tG$ depends on whether $\bk<k<k_>$ or $k>k_>$.  In this case, we set $\Delta\tG = \Delta\tG_{\bk}$ for $\bk_k<k_>$, and $\Delta\tG = \Delta\tG_{\bk} + \Delta\tG_>$ for $k>k_>$, so that $\Delta\tG_>$ is the correction required for $k>k_>$. In effect, $\Delta\tG$ has two branch points, one at $\bk$ and one at $k_>$ in this case, and by making this definition we may write
\begin{equation}
\int_{\bk}^\infty\Delta\tG\ex^{-\mi kx}\,\intd k
= \int_{\bk}^\infty\Delta\tG_{\bk}\ex^{-\mi kx}\,\intd k
+ \int_{k_>}^\infty\Delta\tG_>\ex^{-\mi kx}\,\intd k.
\end{equation}
By considering~\eqref{appendix:JumptG} in this case, we find that
\begin{subequations}
\begin{align}
\Delta\tG_{\bk}=&-\frac{\omega-U(r^*)k}{r^*W(r^*)}
\frac{1}{C^-_1\dg_2-\cg_2D_1+2\mi\upi AD_1\dg_2}\frac{A D_1\dg_2\psi^-_1(\rl)\psi^-_2(\rg)}{C_1^-\dg_2-\cg_2D_1}, \\
\Delta\tG_{>}=&\phantom{{}-}\frac{\omega-U(r^*)k}{2\I\pi r^*W(r^*)}
\frac{1}{C^-_1\dg_2-\cg_2D_1+2\mi\upi AD_1\dg_2}\psi^-_1(\rl)\Delta\psi_2(\rg)\notag, \\
=&-\frac{\omega-U(r^*)k}{r^*W(r^*)}
\frac{A \dg_2 \psi^-_1(\rl)\tp_1(\rg)}{C^-_1\dg_2-\cg_2D_1+2\mi\upi AD_1\dg_2}
\end{align}
\end{subequations}

Finally, when we have $1-h<\rl$ we must consider three cases: $\bk<k<k_<$, $k_<<k<k_>$ and $k_><k$. Similarly to the previous case, we consider $\Delta\tG_{\bk}=\Delta\tG$ for $\bk<k<k_<$, and take $\Delta\tG_{<}$ and $\Delta\tG_{>}$ to be correction terms as $k$ crosses $k_<$ and $k_>$ respectively.  This leads to
\begin{subequations}
\begin{align}
\Delta\tG_{\bk}=&-\frac{\omega-U(r^*)k}{2\I\pi r^*W(r^*)}
\frac{1}{C^-_1\dg_2-\cg_2D_1+2\mi\upi AD_1\dg_2} \notag \\
&\quad\times \left[\frac{2\mi\upi A D_1\dg_2\psi^-_1(\rl)\psi^-_2(\rg)}{C_1^-\dg_2-\cg_2D_1}-\Delta\psi^-_1(\rl)\psi_2(\rg)\right]\notag, \\
=&-\frac{\omega-U(r^*)k}{r^*W(r^*)}
\frac{1}{C^-_1\dg_2-\cg_2D_1+2\mi\upi AD_1\dg_2} \frac{ AD_1^2\psi^-_2(\rl)\psi^-_2(\rg)}{C_1^-\dg_2-\cg_2D_1} \\
\Delta\tG_{<}=&-\frac{\omega-U(r^*)k}{2\I\pi r^*W(r^*)}
\frac{1}{C^-_1\dg_2-\cg_2D_1+2\mi\upi AD_1\dg_2}\Delta\psi^-_1(\rl)\psi_2(\rg)\notag, \\
=&-\frac{\omega-U(r^*)k}{r^*W(r^*)}
\frac{A D_1 \tp_1(\rg)\psi^-_2(\rg)}{C^-_1\dg_2-\cg_2D_1+2\mi\upi AD_1\dg_2} \\
\Delta\tG_{>}=&\phantom{-}\frac{\omega-U(r^*)k}{2\I\pi r^*W(r^*)}
\frac{1}{C^-_1\dg_2-\cg_2D_1+2\mi\upi AD_1\dg_2}\psi^-_1(\rl)\Delta\psi_2(\rg)\notag, \\
=&-\frac{\omega-U(r^*)k}{r^*W(r^*)}
\frac{A \dg_2 \psi^-_1(\rl)\tp_1(\rg)}{C^-_1\dg_2-\cg_2D_1+2\mi\upi AD_1\dg_2}
\end{align}
\end{subequations}

\section{Asymptotic behaviours of $\tG$ and $\Delta\tG_{q}$}\label{appendix:decay-rates}

In order to find the residue contribution of the non-modal pole at $k = k_0$, and the decay rates of the steepest descent contours given by integrating $\Delta\tG_q$ along $k = k_q - \mi\xi$, we are required to understand the behaviour of $\tp_1$ and $\tp_2$ at $r = 1-h,r,r_0,$ and $1$ as $k \to\bk,k_r,k_0$, where $\tp_1$ and $\tp_2$ are given in appendix~\ref{appendix:frobenius-full}.

Considering the evaluations at $r,r_0>1-h$ and $1$, it can be noted that we must examine the cases that $\tp_1$ and $\tp_2$ are described as $\tp_{c1}$ and $\tp_{c2}$ respectively (as described in appendix~\ref{appendix:frobeniusc}), or, if we have been required to perform matching, that both are expressed in terms of $\tp_{11}$ and $\tp_{12}$ (given in appendix~\ref{appendix:frobenius1}).
If we examine the limit $k\to\bk$, it will follow that in each case we are required to take the matched solutions $\tp_{11}$ and $\tp_{12}$.

\subsection{Asymptotic behaviour as \texorpdfstring{$k\to\bk$}{k→ω/M}}

Consider first $\tp_{c1}$ and $\tp_{c2}$ for $k$ close to $\bk$ and $r$ close to $1\!-\!h$.  Since $Q=h\sqrt{1-\omega/(Mk)} = O\big((k-\bk)^\frac{1}{2}\big)$, we consider the limit $|Q|\to 0$ and set $r=1-h+RQ$ for $|R|\leq O(1)$.  By considering the recurrence formulae for the Frobenius expansion coefficients $a_n$ and $b_n$ given in equations~\eqref{appendix:an} and~\eqref{appendix:bn} in this limit, and after some algebra, it can be found to leading order that
\begin{subequations}\label{pc1c2leadingorders}\begin{align}
    \tp_{c1}(1\!-\!h+RQ)&=Q^3(R-1)^3\lr{1+\frac{3}{4}(R-1)+\frac{3}{20}(R-1)^2}+O(Q^4), \label{pc1leadingorder}\\
    \tp_{c2}(1\!-\!h+RQ)&=1+O(Q^2\log(Q)), \\
    \tp_{c1}^\prime(1\!-\!h+RQ)&=3Q^2(R-1)^2\lr{1+(R-1)+\frac{1}{4}(R-1)^2}+O(Q^3), \\
    \tp_{c2}^\prime(1\!-\!h+RQ)&=-Q\log(Q)\!\left(\frac{\omega^2}{M^2} + \frac{m^2}{(1\!-\!h)^2}\right)\!(R\!-\!1)^2\!\lr{\!1+(R\!-\!1)+\frac{(R\!-\!1)^2}{4}}\! +O(Q)
\end{align}\end{subequations}

We consider next $\tp_{11}$ and $\tp_{12}$ in the same limit.  By considering the recurrence formulae for the series coefficients $\alpha_n$ and $\beta_n$ given in equations~\eqref{appendix:alphan} and~\eqref{appendix:betan}, it can be found that there are coefficients $\tp_{11}^{(n)}, \tp_{11}^{\prime(n)},\tp_{12}^{(n)}$, and $\tp_{12}^{\prime(n)}$ which are $O(1)$ as $|Q|\to0$ such that 
\begin{subequations}\label{p1112leadingorders}\begin{align}
    \tp_{11}(1-h+RQ)&=\sum_{n=0}^\infty (RQ)^n \tp_{11}^{(n)}, &
    \tp_{12}(1-h+RQ)&=\sum_{n=0}^\infty (RQ)^n \tp_{12}^{(n)},\\
    \tp_{11}^\prime(1-h+RQ)&=\sum_{n=0}^\infty (RQ)^n \tp_{11}^{\prime(n)}, &
    \tp_{12}^\prime(1-h+RQ)&= \sum_{n=0}^\infty (RQ)^n \tp_{12}^{\prime(n)}.
\end{align}\end{subequations}
Note in particular that, as $|Q|\to0$, the coefficients of the $R^5$ term in both $\tp_{11}$ and $\tp_{12}$ tend to zero at least as fast as $Q^5$, where as in $\tp_{c1}$ from~\eqref{pc1leadingorder} the coefficient of $R^5$ tends to zero as $Q^3$.  Hence, if we were to write $\tp_{c1} = A_1\tp_{11} + B_1\tp_{12}$, then at least one of the coefficients $A_1$ and $B_1$ would need to tend to infinity as $Q^{-2}$ or faster as $|Q|\to0$.  We argue that the choice of $\tp_{11}$ and $\tp_{12}$ as two linearly independent solutions about $r=1$ is arbitrary, and so by symmetry between $\tp_{11}$ and $\tp_{12}$ we expect $A_1$ and $B_1$ to be the same order of magnitude in $Q$, therefore forcing that $A_1 = O(Q^{-2})$ and $B_1 = O(Q^{-2})$.  Similarly, since $\tp_{c2}'$ has a coefficient of $R^4$ which scales as $Q\log Q$, if we were to write $\tp_{c2}' = A_2\tp_{11}' + B_2\tp_{12}'$, then at least one of the coefficients $A_2$ and $B_2$ would need to tend to infinity as $Q^{-3}\log Q$ or faster as $|Q|\to0$, and so we argue that $A_2 = O\big(Q^{-3}\log Q\big)$ and $B_2 = O\big(Q^{-3}\log Q\big)$.

Note, however, that by evaluation the Wronskian $\mathcal{W}(\tp_{c1},\tp_{c2};r)=W(r)$ at $r=1$ using~\eqref{appendix:Wr}, and by definition of $\tp_{11}$ and $\tp_{12}$ at $r=1$, considering $\mathcal{W}(\tp_{11},\tp_{12};r)$ at $r=1$ shows that
\begin{equation}
    A_1B_2-A_2B_1=\frac{3(h-Q)^2(h+Q)^2(1\!-\!h+Q)}{4Q^2}=O(Q^{-2}).
\end{equation}
This is smaller than might have been expected from the individual scalings of $A_1$, $B_1$, $A_2$ and $B_2$ given above, but this is expected as, when the critical point $r_c^+$ is approached, the two linearly independent solutions lose their linear independence, and so there is significant cancellation between $A_1B_2$ and $A_2B_1$.

Note also from~\eqref{appendix:Wr} that, as $|Q|\to0$, we have
\begin{equation}\label{WronskianLeadingOrders}
W(r^*)=\begin{cases}
-\dfrac{3Q^2}{4}\!\left(\!1+\dfrac{Q}{1\!-\!h}\right)\! & r_0\leq1-h \\\\ -\dfrac{3(1\!-\!h - r_0)^4}{4r_0}\!\left(\dfrac{(1\!-\!h)}{Q^2}+\dfrac{1}{Q}+O(1)\!\right)\! &r_0>1-h. \end{cases}
\end{equation}
Assuming that $\tp_{11}$ and $\tp_{12}$ are $O(1)$ when $r$ is not close to $1-h$, it follows that
\begin{subequations}\label{Coeeficientleadingorders}\begin{align}
    C_1&=\frac{4\alpha J_{m}^\prime(\alpha(1-h))}{3Q^2}+O(Q^{-1})=O(Q^{-2})\\
    D_1&=J_{m}(\alpha(1-h))+O(Q)=O(1)\\
    \cg_2&=-\frac{4Q^2}{3}\frac{A_2+\frac{i\omega}{Z}B_2}{(1-h+Q)(h-Q)^2(h+Q)^2}= O(Q^{-1}\log(Q)) \\
    \dg_2&=\phantom{-}\frac{4Q^2}{3}\frac{A_1+\frac{i\omega}{Z}B_1}{(1-h+Q)(h-Q)^2(h+Q)^2}=O(1)\\
    \cl_2&=\phantom{{}-}\frac{\pi\mi(1-h)\alpha}{4}\dg_2H_m^{(2)\prime}(\alpha(1-h))=O(1)\\
    \dl_2&=-\frac{\pi\mi(1-h)\alpha}{4}\dg_2H_m^{(1)\prime}(\alpha(1-h))=O(1)
\end{align}\end{subequations}

We can use the above to establish that $\psi_1$ and $\psi_2$ are both order 1 quantities for particular values of $r$:
\begin{align}
    \psi_1(r)&=J_m(\alpha r)=O(1) &\text{for}\qquad &r<1-h;\label{psi1leading} \\
    \psi_2(r)&=\cg_2\tp_1+\dg_2\tp_2=
    \tp_{12}-\frac{\mi\omega}{Z}\tp_{11}=O(1) &\text{for}\qquad &r>1-h.\label{psi2leading}
\end{align}

We also note that $\omega-U(r^*)k=-M(k-\bk) = -\omega Q^2\!/h^2 + O(Q^4) = O(Q^2)$ for $r_0\leq 1-h$ and is $O(1)$ for $r_0>1-h$, and that
\begin{equation}
    A=-\frac{1}{3}\bbk\lr{\frac{1}{Q}-\frac{1}{r_c^+}}-\frac{2m^2}{3r_c^{+3}}=-\frac{1}{3Q}\bbk+O(1).
\end{equation}

\subsubsection{Behaviour of \texorpdfstring{$\tG$}{G} as \texorpdfstring{$k\to\bk$}{k→ω/M}}
\label{appendix:branchpointbk}

We now use the above scalings to consider the branch point of $\tG$ at $k=\bk$, with the aim of showing that $\tG$ does not experience a pole at $k=\bk$ for any value of $r_0$.  Recall from~\eqref{tildeG} that
\begin{equation}\tag{\ref{tildeG}}
    \tG=\frac{\lr{\omega-U(r^*)k}}{2\upi\mi r^*W(r^*)}\frac{\psi_1(\rl)\psi_2(\rg)}{C_1\dg_2-\cg_2D_1}.
\end{equation}
Using the results above, if $k=\bk+\eps\ex^{\mi\theta}$ then for $r_0\leq 1-h$,
\begin{equation}
    \tG\sim\frac{M\eps\ex^{\mi\theta}}{2\upi\mi (1\!-\!h)}\frac{\psi_1(\rl)\psi_2(\rg)}{\alpha J_m'(\alpha(1\!-\!h))\dg_2} = O(\eps).
\end{equation}

If instead $r_0>1-h$, we find that
\begin{equation}
    \tG\sim\frac{-Mh^4\eps^2\ex^{2\mi\theta}\big(M-U(r_0)\big)}{2\upi\mi\omega(1\!-\!h)(1\!-\!h - r_0)^4}\frac{\psi_1(\rl)\psi_2(\rg)}{\alpha J_m'(\alpha(1-h))\dg_2} = O\big(\eps^2\big).
\end{equation}

In particular, in either case there is no pole of $\tG$ at $k=\bk$.  Hence, we have that
\begin{equation}
I_\eps(x) = \frac{-1}{2\upi}\!\!\int_{0}^{2\upi}\! \tG\!\left(\bk + \eps\ex^{\I\theta}\right) \exp\!\left\{-\mi x\left(\bk+\eps\ex^{\I\theta}\right)\!\right\}
\I\eps\ex^{\I\theta}\,\intd\theta \to 0 \qquad\text{as}\quad \eps \to 0.
\end{equation}

\subsubsection{Behaviour of \texorpdfstring{$\Delta\tG_{\bk}$}{ΔG(ω/M)} as \texorpdfstring{$k\to\bk$}{k→ω/M}}

We now substitute all of the above into the equation for $\Delta\tG_{\bk}$ given in equation~\eqref{JumptGbk}.  First of all, we rewrite~\eqref{JumptGbk} exactly as
\begin{subequations}
\begin{gather}
\Delta\tG_{\bk}=\frac{4A\big(\dg_2\psi_1^-(1\!-\!h)\big)^2 f(r)f(r_0)j(r^*)}{3(1-h)Q^3\lr{C_1^-\dg_2-\cg_2D_1+2\mi\upi AD_1\dg_2}\!\lr{C_1^-\dg_2-\cg_2D_1}},
\\
\text{where}\quad
  j(r_0) = -\frac{3}{4}(1-h)Q^3\frac{\omega-U(r_0)k}{r_0W(r_0)}
  \qquad\text{and}\quad
f(r) = \begin{cases}
\dfrac{\psi_1^-(r)}{\psi_1^-(1\!-\!h)} & r < r_0\\\\
\dfrac{D_1\psi_2^-(r)}{\dg_2\psi_1^-(1\!-\!h)} & r > r_0.
\end{cases}
\end{gather}\end{subequations}
  Taking now the leading order terms as $k \to \bk$, we find that
\begin{subequations}\label{equ:appendix:tGbk}
\begin{align}
\Delta\tG_{\bk}&\sim-\frac{\lr{\frac{\omega^2}{M^2} + \frac{m^2}{(1\!-\!h)^2}}}{4(1-h)}
\frac{J_m\big(\alpha(1\!-\!h)\big)^2}{\alpha^2 J_{m}^\prime\big(\alpha(1\!-\!h)\big)^2}
f(r)f(r_0)j(r_0),
\\\notag\\
\text{where }f(r) &= \begin{cases}
\dfrac{J_m(\alpha r)}{J_m\big(\alpha(1\!-\!h)\big)} & r < r_0\\\\
\dfrac{\psi_2^-(r)}{\dg_2} & r > r_0.
\end{cases}
  \\\notag\\
  \text{and }j(r_0) &=
  \begin{cases}
  -\dfrac{\omega}{h^2}Q^3 & r_0 < 1-h\\\\
  \dfrac{\omega(1-U(r_0)/M)}{(r_0 - 1 + h)^4}Q^5 & r_0 > 1-h.
  \end{cases}
  \end{align}\end{subequations}
  Finally, setting $k = \bk - \mi\xi$, so that $Q = (1-\mi)h\sqrt{M\xi/2\omega} + O\big(\xi^{3/2}\big)$ (recalling that $\Real(Q)\geq 0$), we find that $j(r_0)$ may be written to leading order as
  \begin{equation}\label{equ:appendix:j}
  j(r_0) = \begin{cases}
  \dfrac{1+\mi}{\sqrt{2}}\dfrac{hM^{3/2}}{\omega^{1/2}}\xi^{3/2} & r_0 < 1-h\\\\
  -\dfrac{1-\mi}{\sqrt{2}}\dfrac{h^5M^{5/2}}{\omega^{3/2}}\dfrac{1-U(r_0)/M}{(r_0 - 1 + h)^4}\xi^{5/2} & r_0 > 1-h.
  \end{cases}
  \end{equation}

\subsection{Asymptotic behaviour as \texorpdfstring{$k\to k_0$}{k→k0}}
\label{appendix:decay-rates-r0}

We now consider $k\to k_0$ with $r_0>1-h$.  We have that
\begin{align}\label{k0poler0minrcp}
    r_0-r_c^+&=-\frac{\omega h^2 (k-k_0)}{2Mk_0^2Q_0} + O\big((k-k_0)^2\big), &
    &\text{where}\qquad
    Q_0 = h\sqrt{1 - \frac{\omega}{Mk_0}}.
\end{align}
Hence, in this limit, $\tp_1(r_0)$ and $\tp_2(r_0)$ may always be evaluated in terms of $\tp_{c1}$ and $\tp_{c2}$, as we are always eventually within their radius of convergence.  Hence,  in this limit,
\begin{align}
    \tp_1(r_0)&=\lr{\frac{-\omega h^2 (k-k_0)}{2Mk_0^2Q_0}}^{\!3} + O\big((k-k_0)^4\big), & \tp_2(r_0)&=1+O\big((k-k_0)^2\big).
\end{align}
For $r\not=r_0$, the Bessel function, Hankel functions, and $\tp_1$ and $\tp_2$ all behave as $O(1)$ quantities when evaluated at $1-h, r$, and $1$, resulting in $O(1)$ behaviour for $C_1,D_1,\cl_2,\dl_2,\cg_2$ and $\dg_2$. It can be shown that $A = O(1)$ and that
\begin{equation}\label{k0poleWron}
    W(r_0)=-\frac{3h^4\omega^2(k-k_0)^2}{4Q_0^2M^2k_0^{4}}+O\big((k-k_0)^3\big).
\end{equation}

\subsubsection{Behaviour of $\tG$ as $k \to k_0$ and the residue of the non-modal $k_0$ pole}

Substituting all the above into~\eqref{tildeG} (as $k\to k_0$ from above) gives
\begin{equation}
\tG(k) = \frac{-2Mk_0^2(\omega-Mk_0)}{3\upi\mi r_0h^2\omega(k-k_0)}\frac{1}{C_1^+\dg_2-\cg_2D_1}
\begin{cases}
\dg_2\psi_1(r) & r < r_0\\
D_1\psi_2(r) & r > r_0
\end{cases}
+ O(1),
\end{equation}
confirming a pole at $k=k_0$ that gives a residue contribution once integrated around of
\begin{equation}
\label{appendix:equ:R+}
R_0^+(k_0) = \frac{2Mk_0^2(\omega-Mk_0)\ex^{-\mi k_0x}}{3\pi r_0h^2\omega(C_1^+\dg_2-\cg_2D_1)}
\begin{cases}
\dg_2\psi_1(r) & r < r_0\\
D_1\psi_2(r) & r > r_0.
\end{cases}
\end{equation}

\subsubsection{Behaviour of $\Delta\tG_0$ as $k \to k_0$}

Moreover, we may substitute all the above into $\Delta\tG_0$ from equations~\eqref{JumptGkless} and~\eqref{JumptGkgreat} to find the leading order contribution to $\Delta\tG_0$ as $k\to k_0$.  First of all, we find the exact expression for $\Delta\tG_0$ to be (considering only $r_0>1-h$, as otherwise $\Delta\tG_0 \equiv 0$)
\begin{equation}
\Delta\tG_0 = -\frac{\omega - U(r_0)k}{r_0W(r_0)}\frac{A\tp_1(r_0)}{C_1^-\dg_2-\cg_2D_1+2\mi\upi AD_1\dg_2}
\times\begin{cases}
\dg_2\psi_1^-(r) & r_0 > r\\
D_1\psi_2^-(r) & r_0 < r.
\end{cases}
\end{equation}
Using asymptotics above, to leading order we find that
\begin{equation}
\Delta\tG_0 = \frac{A\omega h^2U(r_0)}{6r_0 Mk_0^2(r_0-1+h)}\frac{ (k-k_0)^2}{C_1^-\dg_2-\cg_2D_1+2\upi\mi AD_1\dg_2}
\times\begin{cases}
\dg_2\psi_1^-(r) & r_0 > r\\
D_1\psi_2^-(r) & r_0 < r.
\end{cases}
\end{equation}

\subsection{Asymptotic behaviour as \texorpdfstring{$k\to k_r$}{k→kr}}
\label{appendix:decay-rates-r}

Analogously to the derivation above for $k\to k_0$, we consider here the limit $k\to k_r$.   In this case, the only difference is that both $W(r^*)$ and $(\omega - U(r^*)k)$ remain $O(1)$ quantities whenever $r\not=r^*$, unlike for the limit $k \to k_0$.  Otherwise, the same procedure is applicable, with, in particular,
\begin{align}
    r-r_c^+&=-\frac{\omega h^2 (k-k_r)}{2Mk_r^2Q_r} + O\big((k-k_r)^2\big), &
    &\text{where}\qquad
    Q_r = h\sqrt{1 - \frac{\omega}{Mk_r}},
\end{align}
and similarly
\begin{align}
    \tp_1(r)&=\lr{\frac{-\omega h^2 (k-k_r)}{2Mk_r^2Q_r}}^{\!3} + O\big((k-k_r)^4\big), & \tp_2(r)&=1+O\big((k-k_r)^2\big).
\end{align}

\subsubsection{Behaviour of $\tG_r$ as $k \to k_r$}

Substituting all the above into~\eqref{tildeG} as $k\to k_r$ gives, to leading order,
\begin{equation}
    \tG\sim\frac{\omega-U(r^*)k_r}{2\upi\mi r^*W(r^*)}\frac{1}{C_1\dg_2-\cg_2D_1}
    \begin{cases}
    D_1\psi_2(r_0) & r < r_0 \\
    \dg_2\psi_1(r_0) & r > r_0
    \end{cases}
     = O(1),
\end{equation}
confirming no singular behaviour at $k=k_r$, and in particular no pole at $k=k_r$.

\subsubsection{Behaviour of $\Delta\tG_r$ as $k \to k_r$}

Equation~\eqref{JumptGkless} and~\eqref{JumptGkgreat} for $r > 1-h$ give $\Delta G_r$ as
\begin{equation}
\Delta\tG_r=-\frac{\omega-U(r^*)k}{ r^* W(r^*)}\frac{ A \tp_1(r) }{C_1^-\dg_2-\cg_2D_1+2\mi\upi A D_1\dg_2} \times \begin{cases}
D_1 \psi_2^-(r_0) & r < r_0\\
\dg_2 \psi_1^-(r_0) &  r > r_0
\end{cases}
\end{equation}
Substituting the above asymptotics into this equation gives
\begin{equation}\label{appendix:DeltatGrlimApp}
\Delta\tG_r\sim\frac{A(\omega-U(r^*)k_r)\omega^3 h^6}{8r^* W(r^*)M^3k_r^6(r-1+h)^3 }\frac{(k-k_r)^3}{C_1^-\dg_2-\cg_2D_1+2\mi\upi A D_1\dg_2}\times \begin{cases}
\dg_2 \psi_1^-(r_0) &  r_0 < r\\
D_1 \psi_2^-(r_0) & r_0 > r.
\end{cases}
\end{equation}

\section{Conjecture on the behaviour of an $n$-polynomial flow profile}\label{appendix:n-polynomial}

In this appendix, we give an argument to support the conjectured behaviour of the critical layer contribution for large $x$ for an $n$-polynomial flow profile given by
\begin{equation}
    U(r)=\begin{cases} M & 0\leq r\leq 1-h \\ M\lr{1-\lr{1-\frac{1-r}{h}}^n} & 1-h\leq r\leq 1 \end{cases}
\end{equation}
The three steepest descent contours will be analogous in form to those given in section~\ref{Sect:Inhomog:Jumps}.  Setting $r_C$ to be some solution of $\omega-U(r_C)k=0$, the solutions for small $|r-r_C|$ will take the form
\begin{subequations}\begin{align}
    \tp_1(r)&=(r-r_C)^3+O((r-r_C)^4) \\
    \tp_2(r)&=A\log(r-r_C)\tp_1(r-r_C)+1+O((r-r_C)^2)\\
    \tp_1^\prime(r)&=3(r-r_C)^2+O((r-r_C)^3) \\
    \tp_2^\prime(r)&=b_2(r-r_C)+O((r-r_C)^2)
\end{align}\end{subequations}
for some coefficient $b_2$.
The Wronskian will satisfy
\begin{equation}
\mathcal{W}(\tp_1,\tp_2;r) = W(r)\propto \frac{1}{r}\prod_{\mathclap{\omega-U(r_c)k=0}}(r-r_c)^2.
\end{equation}
For the solutions expanded around the particular critical point $r_C$, we therefore have
\begin{equation}
    W(r)= -3\frac{r_C}{r}\dfrac{\displaystyle\prod_{\omega-U(r_c)k=0}\!\!\!\!\!\!\!\!(r-r_c)^2}{\displaystyle\prod_{{\begin{array}{c} 
   { \scriptstyle  \omega-U(r_c)k=0 }\\ {\scriptstyle  r_c\not=r_C }
    \end{array}}} \!\!\!\!\!\!\!\!\!\!\!(r_C-r_c)^{2}}.
\vspace{-1em}\end{equation}
As $k\to\bk$ we have
\begin{equation}
    A=-\frac{1}{3}\!\lr{\frac{\omega^2}{M^2}+\frac{m^2}{r_C^2}}\!\!\lr{\frac{U^{\prime\prime}(r_C)}{U^\prime(r_C)}-\frac{1}{r_C}}\!-\frac{2m^2}{3r_C^{3}}
    \sim-\frac{1}{3}\!\lr{\frac{\omega^2}{M^2}+\frac{m^2}{(1\!-\!h)^2}}\!\frac{n-1}{r_C-1\!-\!h}+O(1),
\end{equation}
and also that $W(1-h)=O(1-h-r_C)^2$, and that $W(r)=O((1-h-r_C)^{-2(n-1)})$ for $r>1-h$. 

Because of the $W(r)$ scalings and the $\tp_1$ and $\tp_2$ scalings, we also have that $C_1=O\big((1\!-\!h-r_C)^{-2}\big)$ while $D_1,\cg_2,\dg_2=O(1)$.

It then follows that
\begin{subequations}\begin{align}
    C_1\dg_2-D_1\cg_2&=O\big((1\!-\!h-r_C)^{-2}\big) \\
    \text{and}\quad
    \Delta(C_1\dg_2-D_1\cg_2)&= 2\upi\mi AD_1\dg_2 = O\big((1\!-\!h-r_C)^{-1}\big).
\end{align}\end{subequations}
We further know that as $k\to\bk$ we have
\begin{align}
    \omega-U(1-h)k&=M\lr{k-\bk} & &\text{and}& \psi_1(r),\psi_2(r)&=O(1).
\end{align}
Noting also that $(1-h-r_C)=O\big((k-\bk)^{\frac{1}{n}}\big)$, we may predict the behaviour of $I_{\bk}$:
\begin{equation}\begin{aligned}
    \Delta\tG_{\bk}&\backsim\frac{(\omega-U(r^*)k)A}{r^*W(r^*)\lr{C_1\dg_2-\cg_2D_1}\lr{C_1\dg_2-\cg_2D_1+2\mi\pi AD_1\dg_2}} \\ &\backsim\begin{cases} \dfrac{(k-\bk)(1-h-r_C)^{-1}}{(1-h-r_C)^2 (1-h-r_C)^{-4}}\backsim (k-\bk)^{1+\frac{1}{n}} & r_0\leq1-h \\ \\
    \dfrac{(1-h-r_C)^{-1}}{(1-h-r_C)^{-2(n-1)} (1-h-r_C)^{-4}}\backsim (k-\bk)^{2+\frac{1}{n}} & r_0>1-h,\end{cases}
\end{aligned}\end{equation}
and hence we predict that $I_{\bk}$ decays like $x^{-2-\frac{1}{n}}$ for $r_0\leq1-h$ and $x^{-3-\frac{1}{n}}$ for $r_0>1-h$.

In order to do the same for $I_{r}$ and $I_0$, we first note that $(r-r_C)=O(k-k_r)$ as $k\to k_r$, and analogously for $k\to k_0$. Further we have $C_1,D_1,\cg_2,\dg_2=O(1)$ and that $A=O(1)$.  It is noticed that for $r>1-h$, $\tp_1(r)=O((r-r_C)^3)$  while $\psi_1(r_0),\psi_2(r_0)=O(1)$. 
Using the previously given results for $\omega-U(r^*)k$  and noting that $W(r_0)=O((r_0-r_C)^2)$ for $I_0$ only, and otherwise $W(r_0) = O(1)$, then gives us our results that $I_{0}$ decays like $x^{-3}$ while $I_r$ decays like $x^{-4}$, exactly as for the quadratic and linear cases.

The validity of the above conjecture depends on the the assumed scalings for $\tp_{1}(r)$ and $\tp_{2}(r)$ at $r=1-h$, $1-h<r<1$ and $r=1$, in the limits $k\to\bk$, $k\to k_r$ and $k \to k_0\not=k_r$.   Particular attention would be required for $n\geq 6$, where three expansions would be needed to cover the whole domain $r\in[1-h,1]$.   Moreover, the locations of the $k_\pm$ poles have a significant bearing on the overall far-field magnitude of the critical layer, and in particular whether the $k^+$ occurs as a convective instability or is stabilised by the boundary layer thickness.

\vspace{-1.5em}

\bibliographystyle{jfm}
\phantomsection\addcontentsline{toc}{section}{References}
\bibliography{arxiv}
\end{document}